\documentclass[a4paper,fleqn]{cas-dc}

\usepackage[authoryear,longnamesfirst]{natbib}

\def\tsc#1{\csdef{#1}{\textsc{\lowercase{#1}}\xspace}}
\tsc{WGM}
\tsc{QE}

\begin{document}
\let\WriteBookmarks\relax
\def\floatpagepagefraction{1}
\def\textpagefraction{.001}

\shorttitle{Luciano-Saridakis entropic cosmology}
\shortauthors{Leizerovich et al.}

\title[mode=title]{Observational constraints on Luciano-Saridakis entropic cosmology}

\tnotemark[1]
\tnotetext[1]{}

\affiliation[1]{organization={Universidad de Buenos Aires, Facultad de Ciencias Exactas y Naturales, Departamento de F\'{\i}sica},
            city={Buenos Aires},
            country={Argentina}}

\affiliation[2]{organization={CONICET -- Universidad de Buenos Aires, Instituto de F\'{\i}sica de Buenos Aires (IFIBA)},
            city={Buenos Aires},
            country={Argentina}}

\affiliation[3]{organization={Department of Chemistry, Physics and Environmental and Soil Sciences, Escola Polit\`{e}cnica Superior, Universidad de Lleida},
            addressline={Av. Jaume II, 69},
            postcode={25001},
            city={Lleida},
            country={Spain}}

\affiliation[4]{organization={Institute for Astronomy, Astrophysics, Space Applications and Remote Sensing, National Observatory of Athens},
            postcode={15236},
            city={Penteli},
            country={Greece}}

\affiliation[5]{organization={Department of Physics, National Technical University of Athens},
            addressline={Zografou Campus},
            postcode={157 73},
            city={Athens},
            country={Greece}}

\affiliation[6]{organization={Departamento de Matem\'{a}ticas, Universidad Cat\'{o}lica del Norte},
            addressline={Avda. Angamos 0610, Casilla 1280},
            city={Antofagasta},
            country={Chile}}

\affiliation[7]{organization={CAS Key Laboratory for Researches in Galaxies and Cosmology, Department of Astronomy, University of Science and Technology of China},
            city={Hefei, Anhui 230026},
            country={P.R. China}}

\author[1,2]{Mat\'{\i}as Leizerovich}[orcid=0000-0002-6438-2285]
\cormark[1]
\ead{mleize@df.uba.ar}
\credit{Methodology, Software, Data Curation, Writing, Review and Editing}

\author[1,2]{Susana J. Landau}[orcid=0000-0003-2645-9197]
\ead{slandau@df.uba.ar}
\credit{Methodology, Data Curation, Writing, Review and Editing}

\author[3]{Giuseppe Gaetano Luciano}[orcid=0000-0002-5129-848X]
\ead{giuseppegaetano.luciano@udl.cat}
\credit{Conceptualization, Formal Analysis, Writing, Review and Editing}

\author[4,5]{Andreas Papatriantafyllou}
\ead{apapatriantafyllou@mail.ntua.gr}
\credit{Writing}

\author[4,6,7]{Emmanuel N. Saridakis}[orcid=0000-0003-1500-0874]
\ead{msaridak@noa.gr}
\credit{Conceptualization, Formal Analysis, Writing, Review}

\cortext[1]{Corresponding author}

\begin{abstract}
A recently proposed generalized entropy by Luciano and Saridakis extends the standard Boltzmann-Gibbs and Bekenstein-Hawking framework through a microscopically motivated construction involving two independent entropic exponents. When applied within the gravity-thermodynamics correspondence, this entropy leads to a modified cosmological dynamics that can be interpreted as an effective dark energy sector of entropic origin, while recovering $\Lambda$CDM in appropriate limits. In this work, we perform the first observational confrontation of the resulting entropic cosmology at the background level. Focusing on the case $\alpha_\delta=0$, we constrain the model using Cosmic Chronometers, Pantheon$^+$ Type Ia supernovae calibrated with SH0ES, BAO measurements from DESI DR2 and compressed Planck 2018 CMB information. We find that the model yields a statistically robust fit to the combined data sets and can simultaneously satisfy Pantheon$^+$, SH0ES and CMB shift-parameter constraints, unlike $\Lambda$CDM. Although the entropic parameters remain close to their standard values, the $\Lambda$CDM limit is excluded at the $2\sigma$ level within the restricted parameter space considered. These results indicate that the Luciano-Saridakis entropic cosmology offers a viable extension of the standard model with the potential to alleviate the Hubble tension at the background level.
\end{abstract}








\maketitle


\section{Introduction} \label{sec:outline}

A wide range of observational probes have established that the Universe
has experienced periods of accelerated expansion, both during its very early
history and at late cosmological times. There are two main strategies to 
explain this behavior. 

The first  consists in modifying the 
gravitational sector itself,
namely in extending general relativity by introducing additional degrees of
freedom or higher-order contributions in the gravitational action
\citep{CANTATA:2021asi,Capozziello:2011et,Cai:2015emx}. This approach has led to 
a
variety of modified gravity theories, including curvature-based extensions
such as $f(R)$ gravity \citep{Starobinsky:1980te,Nojiri:2010wj} and $f(G)$ 
gravity
\citep{Nojiri:2005jg,DeFelice:2008wz}, higher-dimensional constructions such as
Lovelock gravity \citep{Lovelock:1971yv,Deruelle:1989fj}, and conformal theories
such as Weyl gravity \citep{Mannheim:1988dj,Flanagan:2006ra}.
Alternative
geometric formulations of gravity based on torsion have also been explored, 
including $f(T)$ gravity
\citep{Ben09,Linder:2010py}, $f(T,T_G)$ gravity
\citep{Kofinas:2014owa}, $f(T,B)$ gravity
\citep{Bahamonde:2015zma}, and teleparallel Horndeski-type theories
\citep{Bahamonde:2019shr}. Similarly, within the non-metric formulation of
gravity, one can construct viable extensions such as $f(Q)$ gravity
\citep{BeltranJimenez:2017tkd,Heisenberg:2023lru} and its generalizations
\citep{De:2023xua}. 

On the other hand, a second, conceptually distinct strategy 
retains general relativity as the
fundamental description of gravity, and instead attributes the accelerated
expansion to additional matter components appearing on the right-hand side of
Einstein field equations. In this case, acceleration is driven by new fields
or effective fluids, such as the inflaton field responsible for primordial
inflation, or by a dark energy sector dominating the late-time cosmic dynamics
\citep{Olive:1989nu,Bartolo:2004if,Copeland:2006wr,Cai:2009zp}.

Beyond these two traditional approaches, there exists a compelling and widely
studied conjecture suggesting that gravity itself has a deep thermodynamic
origin \citep{Jacobson:1995ab,Padmanabhan:2003gd,Padmanabhan:2009vy}. Within this
perspective, spacetime dynamics can be interpreted as emerging from
thermodynamic relations, with horizons playing a central role. In cosmological
settings, treating the Universe as a thermodynamic system bounded by the
apparent horizon \citep{Frolov:2002va,Cai:2005ra,Akbar:2006kj,Cai:2006rs} allows
one to derive the Friedmann equations by applying the first law of
thermodynamics to the horizon. Remarkably, this construction is not restricted
to general relativity, but can be extended to a wide class of modified gravity
theories, provided that the appropriate horizon entropy is 
employed~\citep{Cai:2006rs,Akbar:2006er,Paranjape:2006ca,Jamil:2009eb,Cai:2009ph,
Wang:2009zv, Gim:2014nba,Fan:2014ala}.

Motivated by this viewpoint, considerable attention has been devoted to
generalizations of the standard Boltzmann-Gibbs and Bekenstein-Hawking entropy
expressions. Relaxing the assumption of additivity leads to generalized
statistical entropies such as the R\'enyi entropy \citep{Renyi:1961EEE}, the
Tsallis entropy \citep{Tsallis:1987eu,Lyra:1998wz}, and the Sharma-Mittal 
entropy
\citep{sharma1975new}. Additionally, extensions based on relativistic 
statistical mechanics
give rise to the Kaniadakis entropy \citep{Kaniadakis:2002zz,Kaniadakis:2005zk},
while quantum-gravitational considerations at the horizon level motivate the
Barrow entropy \citep{Barrow:2020tzx}. All these constructions recover the
standard entropy as a particular limit of their defining parameters (see also 
\citep{Nojiri:2022dkr} for a unified multi-parametric generalization of
these entropy functionals). Their
cosmological implications have been extensively explored, and have been shown 
to lead to
rich and phenomenologically viable cosmological scenarios
\citep{Lymperis:2018iuz,Saridakis:2020lrg, 
Geng:2019shx,Lymperis:2021qty,Hernandez-Almada:2021rjs,Zamora:2022cqz,
Luciano:2022ely,Luciano:2022knb,Jizba:2022bfz,Dheepika:2022sio,
Luciano:2023zrx,Luciano:2023fyr,Teimoori:2023hpv,Naeem:2023ipg,
Jalalzadeh:2023mzw,Basilakos:2023kvk,Naeem:2023tcu,Coker:2023yxr,
Saavedra:2023rfq,Nakarachinda:2023jko,Jizba:2023fkp,Okcu:2024tnw, 
Luciano:2023bai,
Jalalzadeh:2024qej,Jizba:2024klq,Ebrahimi:2024zrk,Trivedi:2024inb,
Ghaffari:2022skp,
Okcu:2024llu, Karabat:2024trf,Ens:2024zzs,
Tsilioukas:2024seh,Ualikhanova:2024xxe,Shahhoseini:2025sgl,
Lymperis:2025vup,Nojiri:2025gkq}.

In a recent work,  \citet{Luciano:2026ufu} proposed a new generalized entropy
functional that extends the standard holographic scaling by introducing two
independent exponents. Contrary to constructions
that postulate modified area laws directly at the macroscopic level, this
entropy arises from a well-defined microscopic entropic functional and an
associated generalized microstate counting. When applied to systems with
boundaries, it leads to a generalized holographic-like entropy-area relation,
which reduces to the standard Bekenstein-Hawking expression in appropriate
limits. Within the gravity-thermodynamics correspondence, this entropy gives
rise to a modified cosmological dynamics, which can be conveniently interpreted
in terms of an effective dark energy sector of purely entropic origin.

In the present work, we proceed to a detailed observational confrontation of
the cosmological scenario resulting from the  entropy. Using a
combination of late- and early-Universe datasets, including Cosmic
Chronometers, Type Ia supernovae, Baryon Acoustic Oscillations and Cosmic
Microwave Background measurements, we perform a comprehensive statistical
analysis in order to constrain the free parameters of the model. This allows us
to assess the phenomenological viability of the entropic cosmology and to
quantify possible deviations from the standard $\Lambda$CDM paradigm.

The structure of the paper is as follows. In
Sec.~\ref{sec:framework} we briefly review the theoretical framework of the
 entropy and summarize the resulting modified Friedmann
equations. In Sec.~\ref{sec:data} we describe the observational datasets  
employed in our analysis. The 
statistical methodology and the
corresponding
observational constraints  are presented in
Sec.~\ref{sec:results}. Finally, Sec.~\ref{sec:Conclusions} is devoted to 
the
main conclusions and a discussion of future perspectives. Unless otherwise 
specified, we set the Boltzmann constant and speed of light equal to unity.
  
\section{Modified cosmology from  entropy}\label{sec:framework}
  
\subsection{Generalized entropy with two exponents}

In the conventional framework, equilibrium systems are described through the
Boltzmann-Gibbs-Shannon (BGS) entropy
\begin{equation}
\label{SBG}
S_{\text{BGS}}=-\kappa\sum_{i=1}^{W}p_i\ln p_i\,,
\end{equation}
where $p_i$ denotes the probability of the $i$-th microstate among the $W$
accessible ones, and $\kappa$ fixes the entropy units (in holographic
applications one usually takes $\kappa=k_B$) \citep{goldstein2020gibbs}. Despite
its success, the BGS entropy may be inadequate for systems with long-range
interactions, strong correlations, memory effects or persistent non-equilibrium
dynamics, for which the effective microstate scaling differs from the standard
one.

A broad class of generalized entropies can be written as
\citep{hanel2011comprehensive}
\begin{equation}
\label{Sf}
S_f[p]=\sum_i f(p_i)\,,
\end{equation}
for suitable choices of the real-valued function $f$.
Imposing the four Khinchin axioms, namely  continuity, maximality, 
expandability and separability,
    uniquely selects the BGS entropy
\citep{shannon1948claude,Khinchin1957}. Nevertheless, the axiom of separability 
  is commonly
violated in interacting or effectively non-additive systems, and thus its 
relaxation
naturally motivates generalized entropic forms that remain compatible with a
thermodynamic description.

In gravitational and cosmological settings, generalized entropies acquire a
direct motivation since horizon entropy follows a holographic-like scaling,
\begin{equation}
\label{arealaw}
S_{\text{BGS}}\propto \log W \propto L^2\,,
\end{equation}
where $L$ denotes the characteristic size \citep{tHooft:1993dmi,Susskind:1994vu}
and the first proportionality holds in the microcanonical (equiprobable) case. 
This
implies
\begin{equation}
\label{W}
W=g(L)\,\xi^{L^2}\,,\qquad \xi>1\,,
\end{equation}
with $g(L)$ subleading at large $L$. In $(3+1)$ dimensions this scaling renders
the BGS entropy non-extensive, indicating that generalized entropic
descriptions may be more appropriate.

Motivated by these considerations,  \citet{Luciano:2026ufu} 
introduced a
generalized entropy characterized by two independent exponents. Within the
above axiomatic framework, one defines
\begin{eqnarray}
\nonumber
S_{\delta,\epsilon}&=&\eta_\delta \sum_i 
p_i\left(\log\frac{1}{p_i}\right)^\delta
+\eta_\epsilon \sum_i p_i\left(\log\frac{1}{p_i}\right)^\epsilon\\[1mm]
&=&\eta_\delta\left(\log W\right)^\delta+\eta_\epsilon\left(\log 
W\right)^\epsilon\,,
\label{genentropyexpr}
\end{eqnarray}
for equiprobable distributions, with $\delta,\epsilon>0$. This form corresponds
to the generalized microstate scaling
\begin{equation}
\label{microscaling}
W_{\delta,\epsilon}=g(L)\,\xi^{L^{2\delta}}\,\tilde{\xi}^{L^{2\epsilon}}\,,
\qquad
\xi,\tilde{\xi}>1\,,
\end{equation}
and, for systems with a boundary of area $A\sim L^2$, it leads at large $L$ to
the holographic-like entropy
\begin{equation}
\label{Sde}
S_{\delta,\epsilon}=\gamma_\delta A^\delta+\gamma_\epsilon A^\epsilon\,,
\end{equation}
where $\gamma_\delta$ and $\gamma_\epsilon$ are positive constants of
dimensions $[L^{-2\delta}]$ and $[L^{-2\epsilon}]$, respectively. 

We remark that the generalized microstate scaling \eqref{microscaling} can be 
interpreted as the simplest multi-scaling extension of the standard holographic 
growth of states, as commonly encountered in complex systems with hierarchical 
correlations and multifractal structure (see e.g. \citep{Fractal, 
hanel2011comprehensive}).
From this perspective, the exponents
$(\delta,\epsilon)$ quantify the deviations from the standard holographic 
scaling
through two independent contributions.

Moreover, contrary to purely macroscopic
postulates of modified area-laws,  Eq. \eqref{Sde} follows
from the microscopic entropic functional \eqref{genentropyexpr} and the
microstate scaling \eqref{microscaling}, arising from a controlled violation of
separability. Finally, note that the Bekenstein-Hawking entropy is recovered in 
the limiting cases
\begin{eqnarray}
\label{limits}
 &&\delta=1,\ \gamma_\delta=\frac{1}{4\ell_p^2},\ \gamma_\epsilon=0;\notag\\
&&\epsilon=1,\ \gamma_\delta=0,\ \gamma_\epsilon=\frac{1}{4\ell_p^2};\notag\\
&&\delta=\epsilon=1,\ \gamma_\delta=\gamma_\epsilon=\frac{1}{8\ell_p^2},
 \end{eqnarray}
while for $\delta=\epsilon$ the generalized entropy \eqref{Sde} reduces to the
corresponding single-exponent power-law form \citep{Tsallis:2013,Barrow:2020tzx}.

  \subsection{Gravity-thermodynamics application and modified Friedmann 
equations}

The gravity-thermodynamics conjecture provides a unifying viewpoint according 
to which
gravitational dynamics can be understood as emerging from thermodynamic 
relations
\citep{Jacobson:1995ab,Padmanabhan:2003gd,Padmanabhan:2009vy}. In cosmological 
settings,
this correspondence implies that the Friedmann equations governing the 
expansion of the
Universe can be obtained by applying the first law of thermodynamics to a 
causal horizon,
identified with the apparent horizon of a Friedmann-Robertson-Walker (FRW) 
spacetime
\citep{Frolov:2002va,Cai:2005ra,Akbar:2006kj}. Conversely, modifying the entropy 
associated
with the horizon naturally leads to modified cosmological equations.

In the standard case of general relativity combined with the Bekenstein-Hawking 
entropy,
the apparent horizon entropy is proportional to its area, and the application 
of the first
law reproduces the conventional Friedmann equations, with the cosmological 
constant
appearing as an integration constant 
\citep{Cai:2005ra,Akbar:2006kj}. In these lines, in  \citet{Luciano:2026ufu} the 
above 
procedure was performed, however using the generalized entropy  \eqref{Sde}
instead of the  standard one.

We consider a homogeneous and isotropic spacetime described by the FRW metric
\begin{equation}
ds^2=-dt^2+a^2(t)\left(\frac{dr^2}{1-kr^2}+r^2d\Omega^2 \right),
\label{metric}
\end{equation}
  with $a(t)$   the scale factor, and where $k=0,+1,-1$ denotes   flat, 
close and open spatial geometry respectively. Additionally, we consider that 
the Universe is filled  with
  perfect matter and radiation fluids of energy densities $\rho_m$, $\rho_r$ 
and 
pressures $p_m$, $p_r$ respectively, 
satisfying the
standard continuity equations
\begin{eqnarray}
\label{cont}
&&\dot{\rho}_m+3H(\rho_m+p_m)=0\,,\\
&&\dot{\rho}_r+3H(\rho_r+p_r)=0\,.
\end{eqnarray}
Motivated by current observations, we restrict ourselves to the spatially flat 
case, for
which the apparent-horizon radius reduces to $\tilde r_A=1/H$. Moreover, we 
stress that the generalized entropy modifies the microscopic state counting, 
while the geometric definition of the horizon temperature remains unchanged. 
Therefore, consistently with the standard gravity-thermodynamics correspondence, 
we adopt the usual relation between temperature and surface gravity and assume 
local equilibrium between the horizon and the cosmic fluid 
\citep{Padmanabhan:2009vy,Frolov:2002va,Cai:2005ra,Akbar:2006kj,Jamil:2010di,
Izquierdo:2005ku}. For completeness, we note that non-equilibrium 
generalizations have also been considered in the literature \citep{Eling:2006aw}.

The first law of thermodynamics applied to the apparent horizon reads
\begin{equation}
\label{fltherm}
-\delta Q=dE=T_h dS+WdV\,,
\end{equation}
where $E=(\rho_m+\rho_r) V$ is the total  matter and radiation energy inside 
the horizon, $V$ is the enclosed
volume, and $W=(\rho_m+\rho_r-p_m-p_r)/2$ denotes the work density. The crucial 
modification with
respect to the standard case arises from the entropy variation, which is now 
determined
by the generalized entropy \eqref{Sde}. Differentiating \eqref{Sde} with 
respect to the
horizon radius yields
\begin{equation}
dS_{\delta,\epsilon}
=2\!\left[(4\pi)^{\delta}\gamma_\delta\,\delta\,\tilde r_A^{2\delta-1}
+(4\pi)^{\epsilon}\gamma_\epsilon\,\epsilon\,\tilde r_A^{2\epsilon-1}\right]
d\tilde r_A\,.
\end{equation}
Inserting this expression into the first law \eqref{fltherm}, and following the 
same steps
as in the standard derivation \citep{Cai:2005ra,Akbar:2006kj}, one obtains a 
modified first Friedmann equation of the form \citep{Luciano:2026ufu}
\begin{equation}
\label{FMFE}
H^2=\frac{8\pi G}{3}\left(\rho_m+\rho_r+\rho_{DE}\right)\,,
\end{equation}
where the effects of the generalized entropy are entirely encoded in an 
effective dark
energy component of entropic origin. Its energy density reads
\begin{equation}
\label{EfDEd}
\rho_{DE}
=\frac{3}{8\pi G}\!\left[
H^2\!\left(1-\alpha_\delta H^{2(1-\delta)}-\alpha_\epsilon 
H^{2(1-\epsilon)}\right)
+ C
\right],
\end{equation}
with $C$ an integration constant of dimensions $[L^{-2}]$, and where we have 
defined
\begin{equation}
\alpha_\delta\equiv
\frac{(4\pi)^{\delta}G\,\gamma_\delta\,\delta}{(2-\delta)\pi}\,,\qquad
\alpha_\epsilon\equiv\alpha_{\delta\rightarrow\epsilon}\,.
\end{equation}

Differentiating Eq.~\eqref{FMFE} with respect to cosmic time leads to the 
second modified
Friedmann equation
\begin{equation}
\label{SMFE}
\dot H=-4\pi G\left(\rho_m+p_m+\rho_r+p_r+\rho_{DE}+p_{DE}\right),
\end{equation}
where the effective dark energy pressure is given by
\begin{eqnarray}
\nonumber
&&
\!\!\!\!\!\!\!\!\!\!
p_{DE}=-\frac{1}{8\pi G}\Big\{
3C+3H^2\!\left[1-\alpha_\delta H^{2(1-\delta)}
-\alpha_\epsilon H^{2(1-\epsilon)}\right]\\
&&  \!\!\!\!\!\!
+2\dot H\!\left[1-\alpha_\delta(2-\delta)H^{2(1-\delta)}
-\alpha_\epsilon(2-\epsilon)H^{2(1-\epsilon)}\right]\!
\Big\}.
\label{EfDEp}
\end{eqnarray}
Finally, combining \eqref{EfDEd} and \eqref{EfDEp}, the effective 
equation-of-state parameter of the
entropic dark energy sector is obtained as
{\small{\begin{eqnarray}
\label{EoS}
&&
\!\!\!\!\!\!\!\!\!\!\!\!\!\!\!\!
w_{DE}\equiv\frac{p_{DE}}{\rho_{DE}}\nonumber\\
&&\!\!\!\!\!\!\!
=-1-\frac{2\dot H\!\left[1\!-\!\alpha_\delta(2\!-\!\delta)H^{2(1-\delta)}
\!-\!\alpha_\epsilon(2\!-\!\epsilon)H^{2(1-\epsilon)}\right]}
{3 C+3H^2\left(1-\alpha_\delta H^{2(1-\delta)}
-\alpha_\epsilon H^{2(1-\epsilon)}\right)}\,.
\end{eqnarray}}}

In the three limiting cases \eqref{limits}, the generalized entropy reduces to 
the
Bekenstein-Hawking form and the effective dark energy sector collapses to a 
cosmological
constant $\Lambda$, with $\rho_{DE}= 3C/(8\pi G)$, $p_{DE}=-3C/(8\pi G)$ and 
$C=\Lambda/3$, 
thus recovering
the standard Friedmann equations.

However,
in the general case,  
Eqs.~\eqref{FMFE}-\eqref{EoS}
define a modified cosmological dynamics driven by an entropic dark energy 
component,
whose properties depend on the parameters $(\delta,\epsilon)$ and on the 
normalization
constant $C$. These parameters constitute the theoretical input that will be 
constrained
by observational data in the following sections.

The background dynamics of the universe of modified cosmology through  
generalized entropy
\eqref{FMFE}, can be written as
\begin{equation}
\alpha_\delta H^{4-2\delta}+\alpha_\epsilon H^{4-2\epsilon}=H_0^2 
\left[\Omega_{m0} 
(1+z)^3 + \Omega_{r0} (1+z)^4\right] + C ,
\label{eq:LSEC}
\end{equation}
where $\Omega_{m0}\equiv\frac{8\pi G}{3 H_0^2} \rho_{m0}$ and 
$\Omega_{r0}\equiv\frac{8\pi 
G}{3H_0^2} \rho_{r0}$ are the current density parameters of the matter and 
radiation 
fluids respectively. Furthermore, we have adopted the standard scaling 
$\rho_m=\rho_{m0}\left(1+z\right)^3$ and $\rho_r=\rho_{r0}\left(1+z\right)^4$,
where we have inserted the redshift $z=-1+a_0/a$ as the independent 
variable, setting the current scale factor $a_0$ equal to unity. 

It proves convenient to express  Eq. \eqref{eq:LSEC} in a dimensionless form,  
introducing  $E \equiv H/H_0$, obtaining
\begin{equation}
\tilde{\alpha}_\delta E^{p_\delta}+\tilde{\alpha}_\epsilon E^{p_\epsilon}= 
\Omega_{m0} (1+z)^3 + \Omega_{r0} (1+z)^4 + c \, ,
\label{eq:LSEC_dimensionless}
\end{equation}
where we have introduced the dimensionless variables
\begin{align}
\tilde{\alpha}_\delta &= \alpha_\delta H_0^{2(1-\delta)}, \\
\tilde{\alpha}_\epsilon &= \alpha_\epsilon H_0^{2(1-\epsilon)}, \\
p_\delta &= 4-2\delta, \\
p_\epsilon &= 4-2\epsilon, \\
c &= C / H_0^2.
\end{align}
As usual, applying   Eq.~\eqref{eq:LSEC_dimensionless}    at the current 
redshift $z=0$ gives the 
closure relation
\begin{equation}
c = \tilde{\alpha}_\delta + \tilde{\alpha}_\epsilon - \Omega_{m0} - 
\Omega_{r0} \,,
\label{eq:closure}
\end{equation}
which allows  $c$ to be eliminated in favor of the  other 
quantities.

Equation~\eqref{eq:LSEC_dimensionless} defines an implicit relation for $E(z)$. 
In the special case $\alpha_\delta = 0$, it becomes analytically solvable, 
namely
\begin{equation}
E(z)=\left\{
1+\frac{
\Omega_{m0}[(1+z)^3-1]
+\Omega_{r0}[(1+z)^4-1]
}{\tilde{\alpha}_\epsilon}
\right\}^{\frac{1}{p_\epsilon}}\,,
\label{Hz}
\end{equation}
while the same expression, with the 
substitution $\epsilon \to \delta$, holds for the other special case, namely 
$\alpha_\epsilon = 0$. Also, the dark energy behaves as 
\begin{equation}
  \rho_{DE}(z)=\frac{3 H_0^2}{8 \pi G}E^2(z)-[\rho_m (z) + \rho_r(z)]\,.  
\end{equation}

Equation~\eqref{eq:LSEC_dimensionless} shows strong degeneracies between 
$\alpha_\epsilon$ and 
$\alpha_\delta$, as well as between $\epsilon$ and $\delta$. 
These degeneracies severely compromise the convergence properties of our 
statistical inference framework. We therefore restrict the parameter space by 
setting $\alpha_\delta=0$ in our observational analysis. 
In this way, Eq. \eqref{Hz} encodes the behavior of the background dynamics of 
the 
cosmological model tested against the data in this paper. In the next section, 
the relation between the background dynamics and the  observational quantities 
will be detailed.

\section{Observational data}\label{sec:data}
In this section, we describe the datasets we use to test the  entropic cosmology 
(LSEC) model. Moreover, we discuss some particularities of this model that allow 
to test it with the CMB shift parameters.   

We use data from type Ia supernovae (SNIa), Cosmic Chronometers (CC) , Baryon 
Acoustic Oscillations (BAO), and the Cosmic Microwave Background (CMB) shift 
parameters. Next, we briefly describe each of the considered datasets and their 
connections with the theoretical predictions. In the case of quantities that are 
related to the background dynamics of the universe, the observables are derived 
quantities from the Hubble parameter $H(z)$ or the comoving distance:
\begin{equation}
    \chi(z) = c \int_0^z \frac{dz'}{H(z')}\,,
\end{equation} 
where we have explicitly restored the speed of light $c$. In the following, we 
restrict ourselves to a flat universe and therefore, the latter quantity is 
equal to the transverse comoving distance $D_M(z)$.

The cosmic chronometer (CC) method 
\citep{Moresco:2012by,Moresco:2015cya,Moresco:2016mzx}, enables the determination 
of the Hubble parameter $H(z)$ at different redshifts through differential age 
analysis of passively-evolving elliptical galaxies. This method assumes that 
these galaxies formed synchronously and evolve passively. In this work we 
consider the data set described  in \citet{Chantada2023} (see  Table IX). In 
addition, the full covariance matrix is taken into account, including 
non-diagonal correlations as described in  \citet{2020ApJ...898...82M}.

SNIa are among the most luminous transient events in the Universe. Their 
homogeneous spectral features and standardizable light curves make them optimal 
cosmological probes for distance determination and the estimation of 
cosmological parameters. In this work we use the complete Pantheon$^+$ 
compilation\footnote{Available in https://github .com /PantheonPlusSH0ES 
/DataRelease.} which comprises 1701 data distributed across the redshift range 
$0.001 < z < 2.3$. The compilation also incorporates Cepheid-based distance 
modulus measurements from the SH0ES dataset  ($z<0.01$) for calibration purposes 
\citep{Scolnic:2021amr,Brout:2022vxf} and it is therefore named as PPS 
(Pantheon$^+$ + SH0ES). The distance modulus of SNIa is given by 
\begin{equation}
\mu(z) = 5 \log_{10} \left(\frac{D_L(z)}{\text{Mpc}}\right) + 25,
\end{equation}where $D_L(z)$ is the luminosity distance:
\begin{equation}
    D_L(z) = (1+z) D_M(z).
\end{equation}

Baryon acoustic oscillations (BAO) in the  photon-baryon plasma imprint a 
characteristic scale in the matter distribution at large scales. This scale, 
which corresponds to the sound horizon at the drag epoch $r_d$, represents the 
maximum comoving distance traveled by acoustic waves prior to decoupling. 
Consequently, BAO features serve as a standard ruler for cosmological distance 
measurements. 

Various tracers of the underlying matter density field enable independent 
distance determinations across different redshift ranges. In this work we use 
the recent data provided by the DESI DR2 release \citep{DESI:2025zgx}. The BAO 
feature appears in both the line-of-sight direction and the transverse direction 
and provides a measurement of $\frac{D_H(z)}{r_d}$, and $\frac{D_M(z)}{r_d}$ 
where
\begin{equation}
    D_H(z) = \frac{c}{H(z)}\,,
\end{equation}
and $r_d = r_s(z_d)$ where $z_d$ is the redshift at the drag epoch. We recall 
that the sound horizon $r_s(z)$ is given by 
\begin{equation}
r_s(z)=\int_{z}^{\infty} 
\frac{c_s(z')}{H(z')} \, dz' \, ,
\end{equation}
with $c_s(z)$ the sound speed in the photon-baryon fluid.

The so-called CMB shift parameters ($R$, $l_A$) provide a compressed 
representation of the full CMB anisotropy spectrum through quantities computed 
exclusively from background cosmological evolution. Let us recall that 
\begin{equation}
    R = \sqrt{\Omega_{m0}}\,\frac{H_0}{c}\,D_M(z^*) \, , \quad \quad l_A = \pi 
\frac{D_M(z^*)}{r_s(z^*)} \,,
\end{equation}
where $z^*$ is the redshift at recombination. Therefore, by combining these two 
quantities, one can capture essential information regarding both the physical 
size of the sound horizon at the epoch of recombination and the angular diameter 
distance to the last scattering surface. For cosmological constraints, the 
Planck data release  can be adequately characterized 
by these  parameters in conjunction with the baryon density $\omega_b = \Omega_b 
h^2$ inferred from the same data set.  We use the values calculated in 
\citet{2024PhLB..85438717L} with TT, TE, EE + low E data from the Planck 2018 
release \citep{Planck:2019nip}.

\begin{figure}[ht]
    \centering
    
\includegraphics[width=0.45\textwidth]{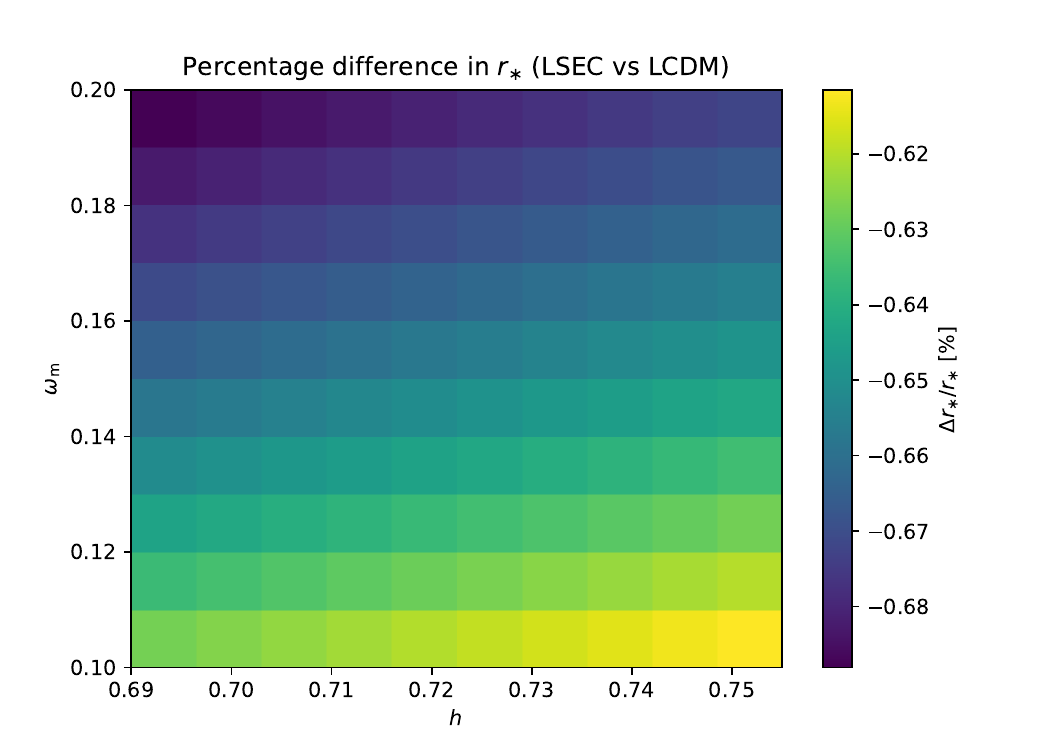}
\\ 
    
\includegraphics[width=0.45\textwidth]{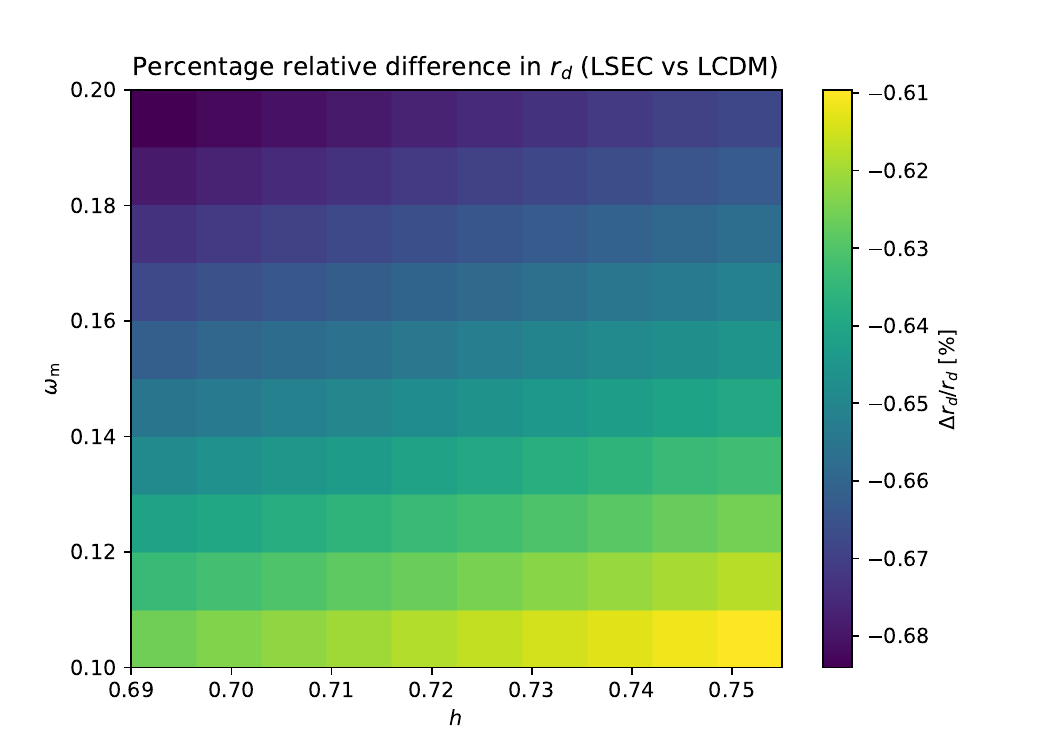}
    \caption{{\it{Percentage difference of the sound horizon calculated in the 
LSEC model with respect to $\Lambda$CDM: Upper panel: Sound horizon at 
recombination, Lower panel: Sound horizon at the drag epoch. The values of $h$, 
$\omega_m$, $\omega_b$, $\tilde{\alpha}_\epsilon$ and $\epsilon$ are fixed to 
their mean values reported in Table \ref{tab:results}.}}}
   \label{fig:soundhorizon}
\end{figure}

The observables discussed above are intrinsically related to three fundamental 
cosmological quantities: the Hubble parameter, the transverse comoving distance, 
and the sound horizon at the recombination/drag epoch. Discrepancies between the 
theoretical predictions of the $\Lambda$CDM model and the LSEC framework emerge 
from differences in the evolution of these quantities. Through preliminary 
analysis, we have determined that the most pronounced differences manifest 
predominantly in the late-time universe. As a result, the sound horizon at the 
recombination/drag epoch shows negligible deviation from its $\Lambda$CDM 
counterpart, as shown in Fig.~\ref{fig:soundhorizon}. This finding carries 
important implications: despite the fact that CMB shift parameters are computed 
under the $\Lambda$CDM assumption, these observables retain their utility for 
constraining alternative cosmological models such as LSEC, since their 
theoretical predictions in this very model deviate less than $1\%$ from 
$\Lambda$CDM, as evidenced in Figs.~\ref{fig:soundhorizon} and \ref{fig:comoving 
distance}.

One might argue that the Planck-derived value of $\Omega_b h^2$ inherently 
assumes $\Lambda$CDM, which is correct. However, since the LSEC model deviates 
significantly from $\Lambda$CDM only in the late universe, while the baryon 
density is primarily determined by early-universe baryon acoustic oscillation 
physics, the $\Lambda$CDM-inferred value should remain a reasonable 
approximation for LSEC.
\begin{figure}[ht]
    \centering
    
\includegraphics[width=0.45\textwidth]{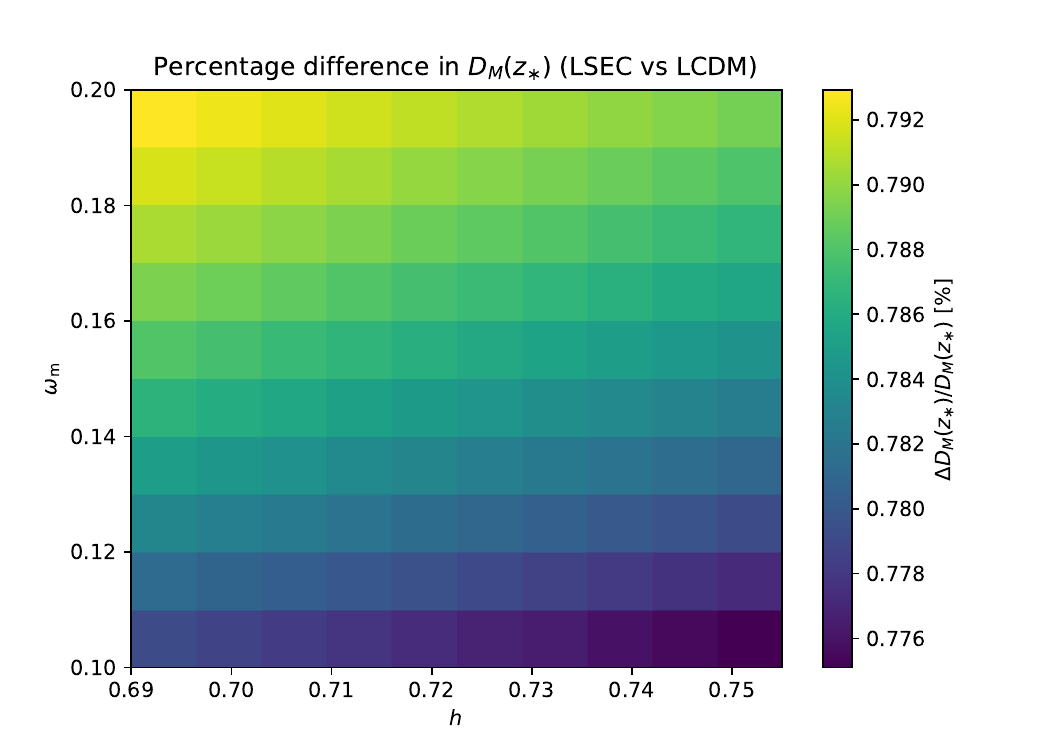}
    \caption{{\it{Percentage difference of the comoving distance calculated in 
the LSEC model with respect to $\Lambda$CDM. The values of $h$, $\omega_m$, 
$\omega_b$, $\tilde{\alpha}_\epsilon$ and $\epsilon$ are fixed to their mean 
values reported in Table~\ref{tab:results}.}}}
   \label{fig:comoving distance}
\end{figure}

\section{Results}\label{sec:results}
We perform a Markov Chain Monte Carlo (MCMC) analysis using the \texttt{COBAYA} 
\citep{Torrado:2020dgo, 2019ascl.soft10019T} sampler with a modified version of 
\texttt{CLASS} Boltzmann code \citep{2011JCAP...07..034B}, varying the 
dimensionless parameters $(h, \omega_m, \omega_b, \tilde{\alpha}_\epsilon, 
\epsilon)$ with appropriate priors, which are shown in Table~\ref{tab:priors}. 
We fix the CMB temperature to $T_{\rm CMB}=2.7255 \, \mathrm{K}$ and assume 
massless neutrinos with an effective number of species $N_{\rm eff}=3.046$.  
Convergence is  assessed using the Gelman-Rubin criterion, requiring $R-1 < 
10^{-2}$, and a burn-in fraction of $30 \%$ is discarded.
\begin{table}[ht]
\centering
\begin{tabular}{|c|c|}
\hline
Parameter & Prior Range \\
\hline
$h$ & $\mathcal{U}[0.4, 1.0]$ \\
$\omega_{m}$ & $\mathcal{U}[0.05, 0.6]$ \\
$\omega_b$ & $\mathcal{U}[0.01, 0.04]$ \\
$\tilde{\alpha}_\epsilon$ & $\mathcal{U}[0.97, 1.12]$ \\
$\epsilon$ & $\mathcal{U}[0.995, 1.015]$ \\
\hline
\end{tabular}
\caption{Priors used in the MCMC analysis for the case $\alpha_\delta=0$. Here, 
$\mathcal{U}[a,b]$ denotes a uniform prior between $a$ and $b$.}
\label{tab:priors}
\end{table}
Before performing the joint analysis of all datasets considered in this work 
(see Sec.~\ref{sec:data}), we verified that the posterior distributions obtained 
from each dataset separately were mutually consistent within the 
Luciano-Saridakis entropic cosmology (LSEC). 

\begin{figure*}
    \centering
    \includegraphics[width=0.7\textwidth]{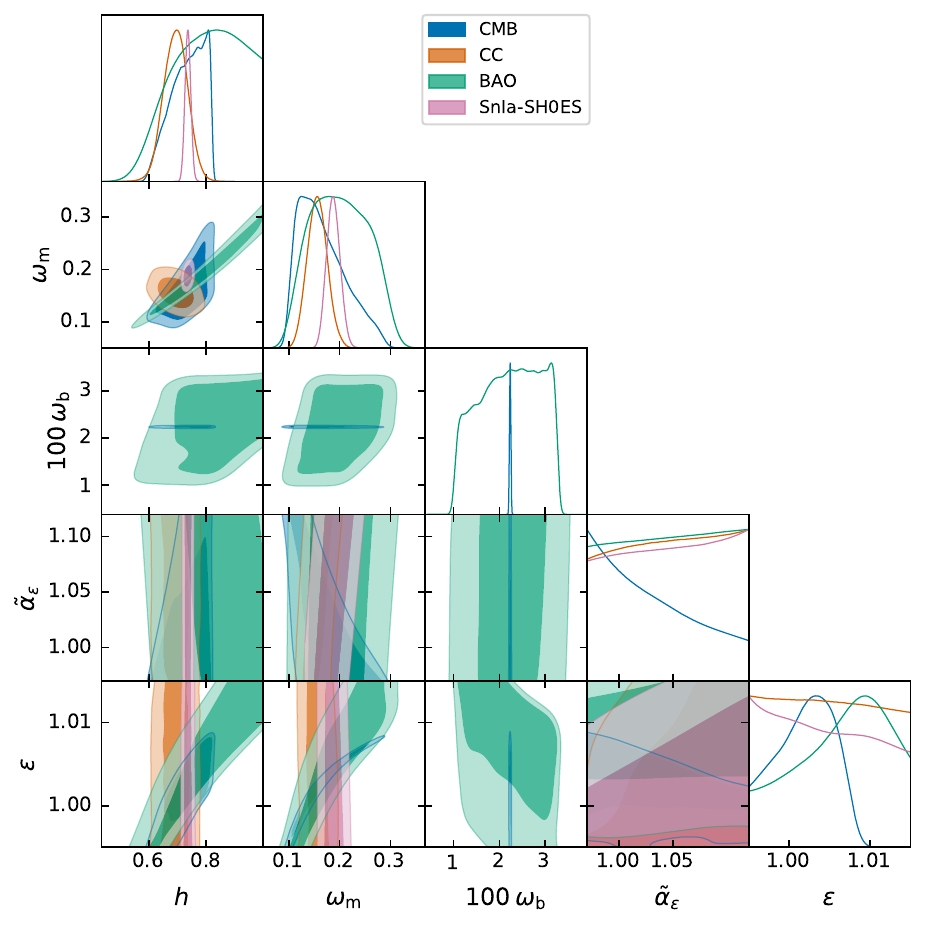}
    \caption{{\it{Results for the statistical analyses considering CMB (Planck 
18 - Shift parameters), CC, SnIa + SH0ES (PPS), BAO (DESI DR2) datasets 
separately. Contour plots and 1-dimensional for the LSEC model with 
$\alpha_\delta=0$. The contours represent 68\% and 95\% confidence levels.}}}
    \label{fig:results_separate_contours_lsec}
\end{figure*}

\begin{figure*}
    \centering
    \includegraphics[width=0.6\textwidth]{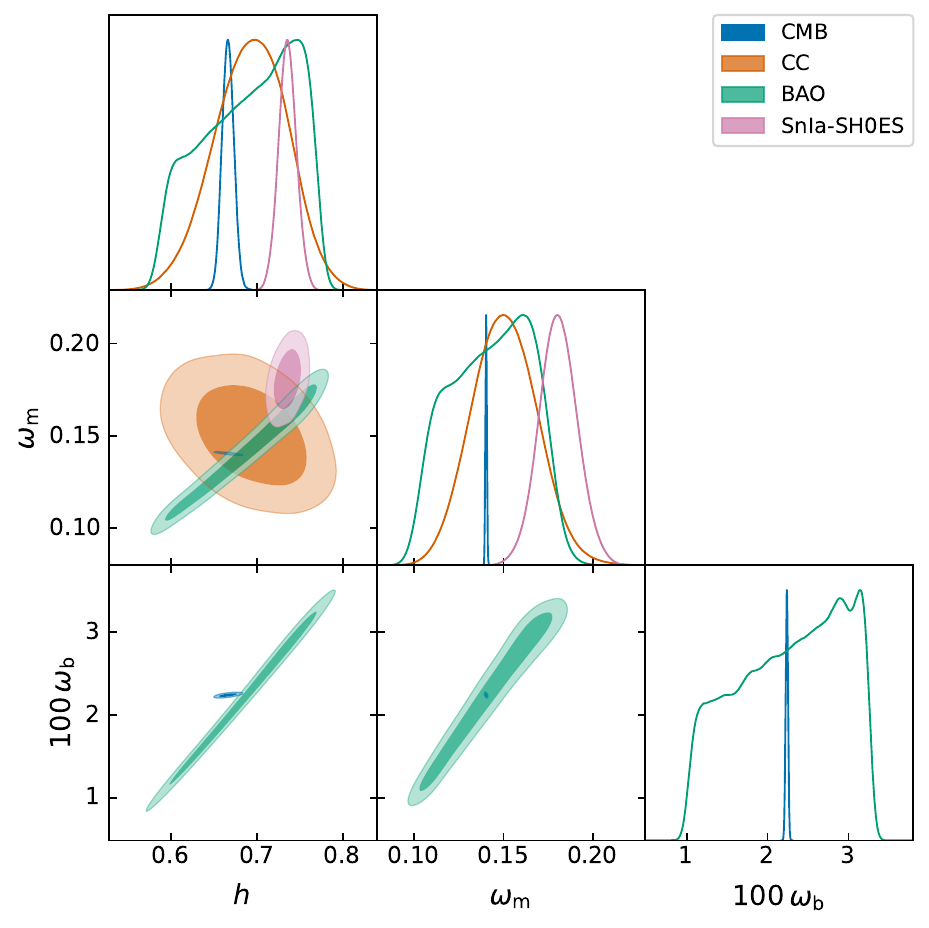}
    \caption{{\it{Results for the statistical analyses considering CMB (Planck 
18 - Shift parameters), CC, SnIa + SH0ES (PPS), BAO (DESI DR2) datasets 
separately. Contour plots and 1-dimensional posteriors for $\rm \Lambda CDM$. 
The contours represent 68\% and 95\% confidence levels.}}}
    \label{fig:results_separate_contours_lcdm}
\end{figure*}

As shown in Fig.~\ref{fig:results_separate_contours_lsec}, the constraints 
derived from PPS and from the CMB shift parameters are compatible in the LSEC 
framework. This behavior contrasts sharply with the $\Lambda$CDM case, where the 
same datasets exhibit a significant inconsistency, as illustrated in 
Fig.~\ref{fig:results_separate_contours_lcdm}. This discrepancy is well known 
and is closely related to the Hubble tension \citep{CANTATA:2021asi}. Therefore, 
an important outcome of the present analysis is that the LSEC model (restricted 
here to $\alpha_\delta=0$) is capable of restoring consistency between PPS and 
compressed CMB information.\footnote{A full resolution of the Hubble tension 
requires consistency between PPS and the complete CMB anisotropy likelihood. 
This lies beyond the scope of the present work and requires the development 
of the perturbative sector of the LSEC model.}

\begin{figure*}
    \centering
    \includegraphics[width=0.7\textwidth]{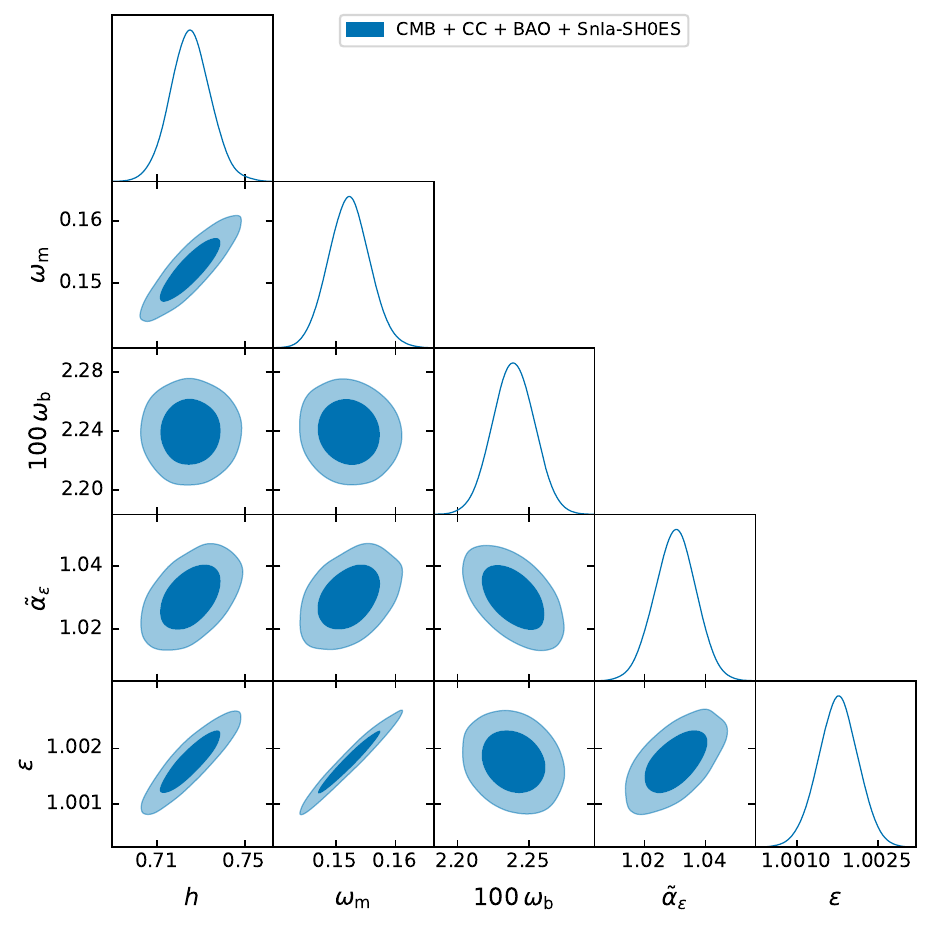}
    \caption{{\it{Posterior distributions for the parameters of the LSEC model 
with 
$\alpha_\delta=0$ using CMB+CC+BAO+SnIa datasets. The contours represent 68\% 
and 95\% confidence levels.}}}
   \label{fig:results}
\end{figure*}

The results of the joint analysis using the CC + PPS + BAO + CMB datasets are 
presented in Fig.~\ref{fig:results} and summarized in Table~\ref{tab:results}. 
The inferred values of the generalized entropy parameters 
$(\alpha_\epsilon,\epsilon)$ lie close to their $\Lambda$CDM limits, 
$\alpha_\epsilon=1$ and $\epsilon=1$, indicating that only mild deviations from 
standard cosmology are required. Nevertheless, within the restricted parameter 
space explored here, the $\Lambda$CDM limit is excluded at the $95\%$ confidence 
level, pointing to a statistically meaningful preference for entropic 
corrections to the late-time cosmological dynamics.
\begin{table}
\centering
\begin{tabular}{lcc}
\hline
Parameter & 68\% C.I. & 95\% C.I. \\
\hline
$h$ & $0.7252\pm 0.0093$ & $0.725^{+0.018}_{-0.018}$\\
$\omega_\mathrm{m}$ & $0.1522\pm 0.0035$ & $0.1522^{+0.0070}_{-0.0067}$\\
$100\,\omega_\mathrm{b}$  & $2.239\pm 0.01$ & $2.239^{+0.029}_{-0.028}$\\
$\tilde{\alpha}_\epsilon$ & $1.0301\pm 0.0068$ & $1.030^{+0.013}_{-0.013}$\\
$\epsilon$ & $1.00176\pm 0.00038$ & $1.00176^{+0.00073}_{-0.00076}$\\
\hline
$\Omega_\mathrm{m0}$ & $0.2895\pm 0.0037$ & $0.2895^{+0.0074}_{-0.0071}$\\
$\alpha_\epsilon$ &  $1.0458\pm 0.0094$ & $1.046^{+0.018}_{-0.019}$\\
\hline
\end{tabular}
\caption{Parameter constraints for the LSEC model with $\alpha_\delta=0$ from 
the joint CMB+CC+BAO+SnIa analysis, showing the 68\% and 95\% intervals.}
\label{tab:results}
\end{table}

Regarding correlations between cosmological and entropic parameters, the 
strongest degeneracy is found between the Hubble constant $h$ and the exponent 
$\epsilon$, as clearly visible in Fig.~\ref{fig:results}. Additional 
correlations involve $\alpha_\epsilon$ with $h$, $\Omega_m$, $\omega_b$, and 
$\epsilon$. These degeneracies originate from the fact that the generalized 
entropy parameters modify the effective evolution of the Hubble rate, allowing 
changes in the late-time expansion history to be compensated by shifts in the 
standard cosmological parameters.

Importantly, these parameter correlations are not merely a statistical artifact 
but reflect a fundamental characteristic of the LSEC model. The entropic 
framework inherently possesses additional degrees of freedom through its extra 
parameters, which naturally endow the model with enhanced flexibility. However, 
it is crucial to emphasize that the mere presence of additional parameters does 
not guarantee improved fits across observational datasets. It is well known that 
many extended cosmological models with extra parameters fail to consistently 
accommodate multiple data groups simultaneously \citep{Frusciante}. 

The distinguishing feature of the LSEC model is its ability to successfully fit 
all observational datasets within a unified framework, effectively reconciling 
PPS distance measurements and CMB shift parameters without introducing 
significant tensions. This capability contrasts with the $\Lambda$CDM model.  In 
this context, the entropic corrections provide an effective physical mechanism 
that relaxes the background-level constraints linking early- and late-Universe 
observables, enabling a more consistent cosmological description.

Notably, the combined analysis rules out values of $\epsilon$ compatible with $\Lambda$CDM within the LSEC framework that are not excluded by any individual data set alone. This results from the interplay between two factors: the correlations between the LSEC model parameters and the cosmological parameters, and the tighter constraints - reflected in the reduced area of the 2D contour plots - obtained when all data sets are combined.


We note that within the $\Lambda$CDM framework the inferred matter density 
parameter $\Omega_m$ exhibits a $2.4\sigma$ tension with the LSEC-inferred value when constrained by PPS data 
alone, increasing to a $3.2\sigma$ discrepancy when using Planck CMB data. 
Similarly, although the inferred value of the Hubble constant in the LSEC model 
is fully consistent with PPS determinations within $\Lambda$CDM \citep{Brout:2022vxf}, it remains in 
strong ($4.8\sigma$) tension with the Planck-derived value when interpreted 
within $\Lambda$CDM \citep{Planck:2018vyg}. These results highlight that the 
Luciano-Saridakis entropic cosmology provides a viable and physically 
well-motivated extension of the standard model, capable of alleviating 
long-standing inconsistencies at the background level while remaining close to 
$\Lambda$CDM in the appropriate limits.

We do not perform model selection criteria such as the Akaike Information Criterion (AIC) or the Bayesian Information Criterion (BIC), as these require a consistent comparison of competing models against the same data sets. Since the SH0ES and CMB data sets are in tension within the $\Lambda$CDM framework, such a comparison is not viable.

Finally, we note that the background dynamics of the model tested against 
observational data in the present work shares the same functional form as that 
analyzed in \citet{Sheykhi:2018dpn,Lymperis:2018iuz}. Furthermore, from an 
observational perspective, a related phenomenological comparison has been 
performed in \citet{Luciano:2025hjn}. However, 
two important differences should be emphasized. First, in the latter study 
one parameter was fixed, corresponding in our framework to the choice 
$\alpha_\epsilon=1$. Second, 
our analysis includes both the SH0ES data set and the CMB shift parameters - 
quantities that are particularly relevant in the context of the Hubble tension - 
whereas the previous work considered only CC, SNIa, and BAO data.

\section{Conclusions}\label{sec:Conclusions}

In this work we have investigated the cosmological implications of the
Luciano-Saridakis generalized entropy, a recently proposed extension of the
standard Boltzmann-Gibbs and Bekenstein-Hawking entropy expressions 
\citep{Luciano:2026ufu}.
Constructed from a well-defined microscopic entropic functional and its 
associated generalized microstate scaling, this entropy, when implemented within 
the gravity-thermodynamics correspondence, gives rise to modified Friedmann 
equations that can be interpreted as describing an effective dark energy 
component of purely entropic origin. Importantly, the standard $\Lambda$CDM 
cosmology is recovered
in appropriate limiting cases, ensuring consistency with conventional
cosmological dynamics while allowing for controlled deviations at late times.

Focusing on the background cosmological evolution, we performed a detailed
statistical analysis of the Luciano-Saridakis entropic cosmology (LSEC) in the
restricted case $\alpha_\delta=0$, employing a combination of Cosmic
Chronometers, SNIa from the Pantheon$^+$ compilation calibrated with SH0ES data, 
BAO measurents from DESI DR2 and compressed CMB information from Planck through 
the shift parameters. We demonstrated
that, within this framework, the different datasets are mutually consistent,
in contrast to the $\Lambda$CDM scenario, where a significant tension persists
between PPS and CMB-derived constraints. This result highlights that the LSEC
model possesses the necessary flexibility to reconcile early- and late-Universe
distance indicators at the background level.

The joint analysis reveals that the best-fit values of the generalized entropy
parameters lie close to, but are statistically distinguishable from, their
$\Lambda$CDM counterparts. Although the deviations are modest, the
$\Lambda$CDM limit is excluded at the $95\%$ confidence level within the
restricted parameter space considered. The presence of characteristic
degeneracies, most notably between the Hubble constant and the entropic
exponent $\epsilon$, plays a central physical role by allowing modifications of
the late-time expansion history that are capable of accommodating both PPS and
CMB shift parameters simultaneously. As a result, the LSEC model has the
potential to alleviate, and possibly resolve, the Hubble tension at the
background level, while remaining close to the standard cosmological model.

Finally, an important next step is the development of the perturbation sector of the LSEC framework, which will allow for a full confrontation with cosmic microwave background anisotropy data and large-scale structure observables. While the present work focuses on the background cosmological evolution, where the model already demonstrates significant consistency with observational datasets, a complete assessment of any cosmological scenario requires the analysis of perturbations. In particular, the behavior of entropic dark energy at the perturbative level, including its impact on structure formation and CMB anisotropies, constitutes a non-trivial extension that cannot be captured within the background framework alone.

For this reason, such an analysis lies beyond the scope of the present work and will be addressed in a dedicated future study. A statistical analysis employing the complete Planck likelihood, combined with a consistent perturbative treatment, is expected to provide tighter constraints and further assess the phenomenological viability of the model. This extension will also enable a more comprehensive exploration of the parameter space, including the simultaneous variation of $\alpha_\epsilon$ and $\alpha_\delta$, and help clarify the role of entropic corrections beyond the background cosmological level.

\printcredits

\bibliographystyle{cas-model2-names}

\bibliography{references}

@article{Luciano:2026ufu,
    author = "Luciano, G. G. and Saridakis, E. N.",
    title = "{New modified cosmology from a new generalized entropy}",
    eprint = "2602.20004",
    archivePrefix = "arXiv",
    primaryClass = "gr-qc",
    month = "2",
    year = "2026"
}

@article{Fractal,
  title = {Fractal measures and their singularities: The characterization of strange sets},
  author = {Halsey, Thomas C. and Jensen, Mogens H. and Kadanoff, Leo P. and Procaccia, Itamar and Shraiman, Boris I.},
  journal = {Phys. Rev. A},
  volume = {33},
  issue = {2},
  pages = {1141--1151},
  numpages = {0},
  year = {1986},
  month = {Feb},
  publisher = {American Physical Society},
  doi = {10.1103/PhysRevA.33.1141},
}

@article{Luciano:2023bai,
    author = "Luciano, Giuseppe Gaetano and Saridakis, Emmanuel",
    title = "{P {\ensuremath{-}} v criticalities, phase transitions and geometrothermodynamics of charged AdS black holes from Kaniadakis statistics}",
    eprint = "2308.12669",
    archivePrefix = "arXiv",
    primaryClass = "gr-qc",
    doi = "10.1007/JHEP12(2023)114",
    journal = "JHEP",
    volume = "12",
    pages = "114",
    year = "2023"
}

@article{Ghaffari:2022skp,
    author = "Ghaffari, S. and Luciano, Giuseppe Gaetano and Capozziello, S.",
    title = "{Barrow holographic dark energy in the Brans{\textendash}Dicke cosmology}",
    eprint = "2209.00903",
    archivePrefix = "arXiv",
    primaryClass = "gr-qc",
    doi = "10.1140/epjp/s13360-022-03481-1",
    journal = "Eur. Phys. J. Plus",
    volume = "138",
    number = "1",
    pages = "82",
    year = "2023"
}

@article{Padmanabhan:2003gd,
    author = "Padmanabhan, T.",
    title = "{Gravity and the thermodynamics of horizons}",
    eprint = "gr-qc/0311036",
    archivePrefix = "arXiv",
    doi = "10.1016/j.physrep.2004.10.003",
    journal = "Phys. Rept.",
    volume = "406",
    pages = "49--125",
    year = "2005"
}

@article{Cai:2005ra,
    author = "Cai, Rong-Gen and Kim, Sang Pyo",
    title = "{First law of thermodynamics and Friedmann equations of Friedmann-Robertson-Walker universe}",
    eprint = "hep-th/0501055",
    archivePrefix = "arXiv",
    doi = "10.1088/1126-6708/2005/02/050",
    journal = "JHEP",
    volume = "02",
    pages = "050",
    year = "2005"
}

@article{Susskind:1994vu,
    author = "Susskind, Leonard",
    title = "{The World as a hologram}",
    eprint = "hep-th/9409089",
    archivePrefix = "arXiv",
    reportNumber = "SU-ITP-94-33",
    doi = "10.1063/1.531249",
    journal = "J. Math. Phys.",
    volume = "36",
    pages = "6377--6396",
    year = "1995"
}

@article{Ebrahimi:2024zrk,
    author = "Ebrahimi, Esmaeil and Sheykhi, Ahmad",
    title = "{Ghost dark energy in Tsallis and Barrow cosmology}",
    eprint = "2405.13096",
    archivePrefix = "arXiv",
    primaryClass = "gr-qc",
    doi = "10.1016/j.dark.2024.101518",
    journal = "Phys. Dark Univ.",
    volume = "45",
    pages = "101518",
    year = "2024"
}

@article{Lymperis:2018iuz,
    author = "Lymperis, Andreas and Saridakis, Emmanuel N.",
    title = "{Modified cosmology through nonextensive horizon thermodynamics}",
    eprint = "1806.04614",
    archivePrefix = "arXiv",
    primaryClass = "gr-qc",
    doi = "10.1140/epjc/s10052-018-6480-y",
    journal = "Eur. Phys. J. C",
    volume = "78",
    number = "12",
    pages = "993",
    year = "2018"
}

@article{Basilakos:2023kvk,
    author = "Basilakos, Spyros and Lymperis, Andreas and Petronikolou, Maria and Saridakis, Emmanuel N.",
    title = "{Alleviating both $H_0$ and $\sigma _8$ tensions in Tsallis cosmology}",
    eprint = "2308.01200",
    archivePrefix = "arXiv",
    primaryClass = "gr-qc",
    doi = "10.1140/epjc/s10052-024-12573-4",
    journal = "Eur. Phys. J. C",
    volume = "84",
    number = "3",
    pages = "297",
    year = "2024"
}

@article{tHooft:1993dmi,
    author = "'t Hooft, Gerard",
    title = "{Dimensional reduction in quantum gravity}",
    eprint = "gr-qc/9310026",
    archivePrefix = "arXiv",
    reportNumber = "THU-93-26",
    journal = "Conf. Proc. C",
    volume = "930308",
    pages = "284--296",
    year = "1993"
}

@incollection{goldstein2020gibbs,
  title={Gibbs and Boltzmann entropy in classical and quantum mechanics},
  author={Goldstein, Sheldon and Lebowitz, Joel L and Tumulka, Roderich and Zangh{\`\i}, Nino},
  booktitle={Statistical mechanics and scientific explanation: Determinism, indeterminism and laws of nature},
  pages={519--581},
  year={2020},
  publisher={World Scientific}
}

@article{Kaniadakis:2002zz,
    author = "Kaniadakis, G.",
    title = "{Statistical mechanics in the context of special relativity}",
    eprint = "cond-mat/0210467",
    archivePrefix = "arXiv",
    doi = "10.1103/PhysRevE.66.056125",
    journal = "Phys. Rev. E",
    volume = "66",
    pages = "056125",
    year = "2002"
}

@article{Tsallis:1987eu,
    author = "Tsallis, Constantino",
    title = "{Possible Generalization of Boltzmann-Gibbs Statistics}",
    reportNumber = "CBPF-NF-062-87",
    doi = "10.1007/BF01016429",
    journal = "J. Statist. Phys.",
    volume = "52",
    pages = "479--487",
    year = "1988"
}

@book{Khinchin1957,
  author    = {A.I. Khinchin},
  title     = {Mathematical Foundations of Information Theory},
  publisher = {Dover Publications},
  year      = {1957},
  address   = {New York, NY, USA}
}

@article{Tsallis:2013,
    author = "Tsallis, Constantino and Cirto, Leonardo J. L.",
    title = "{Black hole thermodynamical entropy}",
    doi = "10.1140/epjc/s10052-013-2487-6",
    journal = "Eur. Phys. J. C",
    volume = "73",
    number = "7",
    year = "2013"
}

@article{Eling:2006aw,
    author = "Eling, Christopher and Guedens, Raf and Jacobson, Ted",
    title = "{Non-equilibrium thermodynamics of spacetime}",
    eprint = "gr-qc/0602001",
    archivePrefix = "arXiv",
    doi = "10.1103/PhysRevLett.96.121301",
    journal = "Phys. Rev. Lett.",
    volume = "96",
    pages = "121301",
    year = "2006"
}

@article{Barrow:2020tzx,
    author = "Barrow, John D.",
    title = "{The Area of a Rough Black Hole}",
    eprint = "2004.09444",
    archivePrefix = "arXiv",
    primaryClass = "gr-qc",
    doi = "10.1016/j.physletb.2020.135643",
    journal = "Phys. Lett. B",
    volume = "808",
    pages = "135643",
    year = "2020"
}

@article{shannon1948claude,
  title={Claude Elwood Shannon},
  author={Shannon, Claude Elwood},
  journal={Bell Syst. Tech. J},
  volume={27},
  pages={379--423},
  year={1948}
}

@article{hanel2011comprehensive,
  title={A comprehensive classification of complex statistical systems and an axiomatic derivation of their entropy and distribution functions},
  author={Hanel, Rudolf and Thurner, Stefan},
  journal={Europhysics Letters},
  volume={93},
  number={2},
  pages={20006},
  year={2011},
  publisher={IOP Publishing}
}

@article{Jizba:2024klq,
    author = "Jizba, Petr and Lambiase, Gaetano and Luciano, Giuseppe Gaetano and Mastrototaro, Leonardo",
    title = "{Imprints of Barrow\textendash{}Tsallis cosmology in primordial gravitational waves}",
    eprint = "2403.09797",
    archivePrefix = "arXiv",
    primaryClass = "gr-qc",
    doi = "10.1140/epjc/s10052-024-13455-5",
    journal = "Eur. Phys. J. C",
    volume = "84",
    number = "10",
    pages = "1076",
    year = "2024"
}

@article{Jacobson:1995ab,
    author = "Jacobson, Ted",
    title = "{Thermodynamics of space-time: The Einstein equation of state}",
    eprint = "gr-qc/9504004",
    archivePrefix = "arXiv",
    reportNumber = "UMDGR-95-114",
    doi = "10.1103/PhysRevLett.75.1260",
    journal = "Phys. Rev. Lett.",
    volume = "75",
    pages = "1260--1263",
    year = "1995"
}

@article{Padmanabhan:2009vy,
    author = "Padmanabhan, T.",
    title = "{Thermodynamical Aspects of Gravity: New insights}",
    eprint = "0911.5004",
    archivePrefix = "arXiv",
    primaryClass = "gr-qc",
    doi = "10.1088/0034-4885/73/4/046901",
    journal = "Rept. Prog. Phys.",
    volume = "73",
    pages = "046901",
    year = "2010"
}

@article{Frolov:2002va,
    author = "Frolov, Andrei V. and Kofman, Lev",
    title = "{Inflation and de Sitter thermodynamics}",
    eprint = "hep-th/0212327",
    archivePrefix = "arXiv",
    reportNumber = "CITA-2002-46",
    doi = "10.1088/1475-7516/2003/05/009",
    journal = "JCAP",
    volume = "05",
    pages = "009",
    year = "2003"
}

@article{Akbar:2006kj,
    author = "Akbar, M. and Cai, Rong-Gen",
    title = "{Thermodynamic Behavior of Friedmann Equations at Apparent Horizon of FRW Universe}",
    eprint = "hep-th/0609128",
    archivePrefix = "arXiv",
    doi = "10.1103/PhysRevD.75.084003",
    journal = "Phys. Rev. D",
    volume = "75",
    pages = "084003",
    year = "2007"
}

@article{Paranjape:2006ca,
    author = "Paranjape, Aseem and Sarkar, Sudipta and Padmanabhan, T.",
    title = "{Thermodynamic route to field equations in Lancos-Lovelock gravity}",
    eprint = "hep-th/0607240",
    archivePrefix = "arXiv",
    doi = "10.1103/PhysRevD.74.104015",
    journal = "Phys. Rev. D",
    volume = "74",
    pages = "104015",
    year = "2006"
}

@article{Akbar:2006er,
    author = "Akbar, M. and Cai, Rong-Gen",
    title = "{Friedmann equations of FRW universe in scalar-tensor gravity, f(R) gravity and first law of thermodynamics}",
    eprint = "hep-th/0602156",
    archivePrefix = "arXiv",
    doi = "10.1016/j.physletb.2006.02.035",
    journal = "Phys. Lett. B",
    volume = "635",
    pages = "7--10",
    year = "2006"
}

@article{Jamil:2009eb,
    author = "Jamil, Mubasher and Saridakis, Emmanuel N. and Setare, M. R.",
    title = "{Thermodynamics of dark energy interacting with dark matter and radiation}",
    eprint = "0910.0822",
    archivePrefix = "arXiv",
    primaryClass = "hep-th",
    doi = "10.1103/PhysRevD.81.023007",
    journal = "Phys. Rev. D",
    volume = "81",
    pages = "023007",
    year = "2010"
}

@article{Cai:2009ph,
    author = "Cai, Rong-Gen and Ohta, Nobuyoshi",
    title = "{Horizon Thermodynamics and Gravitational Field Equations in Horava-Lifshitz Gravity}",
    eprint = "0910.2307",
    archivePrefix = "arXiv",
    primaryClass = "hep-th",
    reportNumber = "KU-TP-035",
    doi = "10.1103/PhysRevD.81.084061",
    journal = "Phys. Rev. D",
    volume = "81",
    pages = "084061",
    year = "2010"
}

@article{Wang:2009zv,
    author = "Wang, Mengjie and Jing, Jiliang and Ding, Chikun and Chen, Songbai",
    title = "{First law of thermodynamics in IR modified Ho\v{r}ava-Lifshitz gravity}",
    eprint = "0912.4832",
    archivePrefix = "arXiv",
    primaryClass = "gr-qc",
    doi = "10.1103/PhysRevD.81.083006",
    journal = "Phys. Rev. D",
    volume = "81",
    pages = "083006",
    year = "2010"
}

@article{Jamil:2010di,
    author = "Jamil, Mubasher and Saridakis, Emmanuel N. and Setare, M. R.",
    title = "{The generalized second law of thermodynamics in Horava-Lifshitz cosmology}",
    eprint = "1003.0876",
    archivePrefix = "arXiv",
    primaryClass = "hep-th",
    doi = "10.1088/1475-7516/2010/11/032",
    journal = "JCAP",
    volume = "11",
    pages = "032",
    year = "2010"
}

@article{Gim:2014nba,
    author = "Gim, Yongwan and Kim, Wontae and Yi, Sang-Heon",
    title = "{The first law of thermodynamics in Lifshitz black holes revisited}",
    eprint = "1403.4704",
    archivePrefix = "arXiv",
    primaryClass = "hep-th",
    doi = "10.1007/JHEP07(2014)002",
    journal = "JHEP",
    volume = "07",
    pages = "002",
    year = "2014"
}

@article{Fan:2014ala,
    author = "Fan, Zhong-Ying and Lu, H.",
    title = "{Thermodynamical First Laws of Black Holes in Quadratically-Extended Gravities}",
    eprint = "1501.00006",
    archivePrefix = "arXiv",
    primaryClass = "hep-th",
    doi = "10.1103/PhysRevD.91.064009",
    journal = "Phys. Rev. D",
    volume = "91",
    number = "6",
    pages = "064009",
    year = "2015"
}

@article{Luciano:2022knb,
    author = "Luciano, Giuseppe Gaetano",
    title = "{Modified Friedmann equations from Kaniadakis entropy and cosmological implications on baryogenesis and ${}^7 Li$-abundance}",
    doi = "10.1140/epjc/s10052-022-10285-1",
    journal = "Eur. Phys. J. C",
    volume = "82",
    number = "4",
    pages = "314",
    year = "2022"
}

@article{Sheykhi:2018dpn,
    author = "Sheykhi, Ahmad",
    title = "{Modified Friedmann Equations from Tsallis Entropy}",
    eprint = "1806.03996",
    archivePrefix = "arXiv",
    primaryClass = "gr-qc",
    doi = "10.1016/j.physletb.2018.08.036",
    journal = "Phys. Lett. B",
    volume = "785",
    pages = "118--126",
    year = "2018"
}

@article{Luciano:2023fyr,
    author = "Luciano, Giuseppe Gaetano and Sheykhi, Ahmad",
    title = "{Black hole geometrothermodynamics and critical phenomena: A look from Tsallis entropy-based perspective}",
    eprint = "2304.11006",
    archivePrefix = "arXiv",
    primaryClass = "hep-th",
    doi = "10.1016/j.dark.2023.101319",
    journal = "Phys. Dark Univ.",
    volume = "42",
    pages = "101319",
    year = "2023"
}

@article{Luciano:2023zrx,
    author = "Luciano, G. G.",
    title = "{From the emergence of cosmic space to horizon thermodynamics in Barrow entropy-based Cosmology}",
    doi = "10.1016/j.physletb.2023.137721",
    journal = "Phys. Lett. B",
    volume = "838",
    pages = "137721",
    year = "2023"
}

@article{Luciano:2022ely,
    author = "Luciano, Giuseppe Gaetano and Gine, Jaume",
    title = "{Baryogenesis in non-extensive Tsallis Cosmology}",
    eprint = "2204.02723",
    archivePrefix = "arXiv",
    primaryClass = "gr-qc",
    doi = "10.1016/j.physletb.2022.137352",
    journal = "Phys. Lett. B",
    volume = "833",
    pages = "137352",
    year = "2022"
}

@article{Planck:2018vyg,
    author = "Aghanim, N. and others",
    collaboration = "Planck",
    title = "{Planck 2018 results. VI. Cosmological parameters}",
    eprint = "1807.06209",
    archivePrefix = "arXiv",
    primaryClass = "astro-ph.CO",
    doi = "10.1051/0004-6361/201833910",
    journal = "Astron. Astrophys.",
    volume = "641",
    pages = "A6",
    year = "2020",
    note = "[Erratum: Astron.Astrophys. 652, C4 (2021)]"
}

@article{Planck:2019nip,
    author = "Aghanim, N. and others",
    collaboration = "Planck",
    title = "{Planck 2018 results. V. CMB power spectra and likelihoods}",
    eprint = "1907.12875",
    archivePrefix = "arXiv",
    primaryClass = "astro-ph.CO",
    doi = "10.1051/0004-6361/201936386",
    journal = "Astron. Astrophys.",
    volume = "641",
    pages = "A5",
    year = "2020"
}

@article{Saridakis:2020lrg,
    author = "Saridakis, Emmanuel N.",
    title = "{Modified cosmology through spacetime thermodynamics and Barrow horizon entropy}",
    eprint = "2006.01105",
    archivePrefix = "arXiv",
    primaryClass = "gr-qc",
    doi = "10.1088/1475-7516/2020/07/031",
    journal = "JCAP",
    volume = "07",
    pages = "031",
    year = "2020"
}

@article{Izquierdo:2005ku,
    author = "Izquierdo, German and Pavon, Diego",
    title = "{Dark energy and the generalized second law}",
    eprint = "astro-ph/0505601",
    archivePrefix = "arXiv",
    doi = "10.1016/j.physletb.2005.12.040",
    journal = "Phys. Lett. B",
    volume = "633",
    pages = "420--426",
    year = "2006"
}

@article{Tsilioukas:2024seh,
    author = "Tsilioukas, Stylianos A. and Petropoulos, Nicholas and Saridakis, Emmanuel N.",
    title = "{Topological dark energy from black-hole formations and mergers through the gravity-thermodynamics approach}",
    eprint = "2412.21146",
    archivePrefix = "arXiv",
    primaryClass = "gr-qc",
    doi = "10.1103/PhysRevD.111.103514",
    journal = "Phys. Rev. D",
    volume = "111",
    number = "10",
    pages = "103514",
    year = "2025"
}

@article{Frusciante,
    author = "Frusciante, Noemi and Perenon, Louis",
    title = "{Effective field theory of dark energy: A review}",
    eprint = "1907.03150",
    archivePrefix = "arXiv",
    primaryClass = "astro-ph.CO",
    doi = "10.1016/j.physrep.2020.02.004",
    journal = "Phys. Rept.",
    volume = "857",
    pages = "1--63",
    year = "2020"
}

@article{Lymperis:2025vup,
    author = "Lymperis, Andreas and Nugmanova, Gulgassyl and Sergazina, Almira",
    title = "{Correspondence between Myrzakulov $F(R,Q)$ gravity and Tsallis 
cosmology}",
    eprint = "2502.04462",
    archivePrefix = "arXiv",
    primaryClass = "gr-qc",
    month = "2",
    year = "2025"
}

@article{Lymperis:2021qty,
    author = "Lymperis, Andreas and Basilakos, Spyros and Saridakis, Emmanuel 
N.",
    title = "{Modified cosmology through Kaniadakis horizon entropy}",
    eprint = "2108.12366",
    archivePrefix = "arXiv",
    primaryClass = "gr-qc",
    doi = "10.1140/epjc/s10052-021-09852-9",
    journal = "Eur. Phys. J. C",
    volume = "81",
    number = "11",
    pages = "1037",
    year = "2021"
}

@article{Hernandez-Almada:2021rjs,
    author = "Hern\'andez-Almada, A. and Leon, Genly and Maga\~na, Juan and 
Garc\'\i{}a-Aspeitia, Miguel A. and Motta, V. and Saridakis, Emmanuel N. and 
Yesmakhanova, Kuralay and Millano, Alfredo D.",
    title = "{Observational constraints and dynamical analysis of Kaniadakis 
horizon-entropy cosmology}",
    eprint = "2112.04615",
    archivePrefix = "arXiv",
    primaryClass = "astro-ph.CO",
    doi = "10.1093/mnras/stac795",
    journal = "Mon. Not. Roy. Astron. Soc.",
    volume = "512",
    number = "4",
    pages = "5122--5134",
    year = "2022"
}

@article{Jizba:2022bfz,
    author = "Jizba, Petr and Lambiase, Gaetano",
    title = "{Tsallis cosmology and its applications in dark matter physics with 
focus on IceCube high-energy neutrino data}",
    eprint = "2206.12910",
    archivePrefix = "arXiv",
    primaryClass = "hep-th",
    doi = "10.1140/epjc/s10052-022-11113-2",
    journal = "Eur. Phys. J. C",
    volume = "82",
    number = "12",
    pages = "1123",
    year = "2022"
}

@article{Jizba:2023fkp,
    author = "Jizba, Petr and Lambiase, Gaetano",
    title = "{Constraints on Tsallis Cosmology from Big Bang Nucleosynthesis and 
the Relic Abundance of Cold Dark Matter Particles}",
    eprint = "2310.19045",
    archivePrefix = "arXiv",
    primaryClass = "gr-qc",
    doi = "10.3390/e25111495",
    journal = "Entropy",
    volume = "25",
    number = "11",
    pages = "1495",
    year = "2023"
}

@article{Dheepika:2022sio,
    author = "Dheepika, M. and T., Hassan Basari V. and Mathew, Titus K.",
    title = "{Emergence of cosmic space in Tsallis modified gravity from 
equilibrium and non-equilibrium thermodynamic perspective}",
    eprint = "2211.14039",
    archivePrefix = "arXiv",
    primaryClass = "gr-qc",
    doi = "10.1088/1402-4896/ad1375",
    journal = "Phys. Scripta",
    volume = "99",
    number = "1",
    pages = "015014",
    year = "2024"
}

@article{Karabat:2024trf,
    author = "Karabat, M. Faruk",
    title = "{Barrow entropic cosmology with exponential potential field}",
    doi = "10.1142/S0217732324501098",
    journal = "Mod. Phys. Lett. A",
    volume = "39",
    number = "23n24",
    pages = "2450109",
    year = "2024"
}

@article{Capozziello:2011et,
    author = "Capozziello, Salvatore and De Laurentis, Mariafelicia",
    title = "{Extended Theories of Gravity}",
    eprint = "1108.6266",
    archivePrefix = "arXiv",
    primaryClass = "gr-qc",
    doi = "10.1016/j.physrep.2011.09.003",
    journal = "Phys. Rept.",
    volume = "509",
    pages = "167--321",
    year = "2011"
}

@book{CANTATA:2021asi,
    author = "Saridakis, Emmanuel N. and others",
    editor = "Saridakis, Emmanuel N. and Lazkoz, Ruth and Salzano, Vincenzo and Vargas Moniz, Paulo and Capozziello, Salvatore and Beltr{\'a}n Jim{\'e}nez, Jose and De Laurentis, Mariafelicia and Olmo, Gonzalo J.",
    collaboration = "CANTATA",
    title = "{Modified Gravity and Cosmology. An Update by the CANTATA Network}",
    eprint = "2105.12582",
    archivePrefix = "arXiv",
    primaryClass = "gr-qc",
    doi = "10.1007/978-3-030-83715-0",
    isbn = "978-3-030-83714-3, 978-3-030-83717-4, 978-3-030-83715-0",
    publisher = "Springer",
    year = "2021"
}

@article{Cai:2015emx,
    author = "Cai, Yi-Fu and Capozziello, Salvatore and De Laurentis, 
Mariafelicia and Saridakis, Emmanuel N.",
    title = "{f(T) teleparallel gravity and cosmology}",
    eprint = "1511.07586",
    archivePrefix = "arXiv",
    primaryClass = "gr-qc",
    doi = "10.1088/0034-4885/79/10/106901",
    journal = "Rept. Prog. Phys.",
    volume = "79",
    number = "10",
    pages = "106901",
    year = "2016"
}

@article{Starobinsky:1980te,
    author = "Starobinsky, A. A.",
    title = "{A New Type of Isotropic Cosmological Models Without Singularity}",
    doi = "10.1016/0370-2693(80)90670-X",
    journal = "Phys. Lett. B",
    volume = "91",
    pages = "99",
    year = "1980"
}

@article{Nojiri:2010wj,
    author = "Nojiri, Shin'ichi and Odintsov, Sergei D.",
    title = "{Unified cosmic history in modified gravity: from F(R) theory to 
Lorentz non-invariant models}",
    eprint = "1011.0544",
    archivePrefix = "arXiv",
    primaryClass = "gr-qc",
    doi = "",
    journal = "Phys. Rept.",
    volume = "505",
    pages = "59",
    year = "2011"
}

@article{Nojiri:2005jg,
    author = "Nojiri, Shin'ichi and Odintsov, Sergei D.",
    title = "{Modified Gauss-Bonnet theory as gravitational alternative for dark 
energy}",
    eprint = "hep-th/0508049",
    archivePrefix = "arXiv",
    primaryClass = "hep-th",
    doi = "10.1016/j.physletb.2005.10.010",
    journal = "Phys. Lett. B",
    volume = "631",
    pages = "1",
    year = "2005"
}

@article{DeFelice:2008wz,
    author = "De Felice, Antonio and Tsujikawa, Shinji",
    title = "{Construction of cosmologically viable f(G) dark energy models}",
    eprint = "0810.5712",
    archivePrefix = "arXiv",
    primaryClass = "hep-th",
    doi = "10.1016/j.physletb.2009.03.063",
    journal = "Phys. Lett. B",
    volume = "675",
    pages = "1",
    year = "2009"
}

@article{Lovelock:1971yv,
    author = "Lovelock, David",
    title = "{The Einstein tensor and its generalizations}",
    doi = "10.1063/1.1665683",
    journal = "J. Math. Phys.",
    volume = "12",
    pages = "498",
    year = "1971"
}

@article{Deruelle:1989fj,
    author = "Deruelle, N. and Farina-Busto, L.",
    title = "{The Lovelock Gravitational Field Equations in Cosmology}",
    doi = "10.1103/PhysRevD.41.3696",
    journal = "Phys. Rev. D",
    volume = "41",
    pages = "3696",
    year = "1990"
}

@article{Mannheim:1988dj,
    author = "Mannheim, Philip D. and Kazanas, Demosthenes",
    title = "{Exact Vacuum Solution to Conformal Weyl Gravity and Galactic 
Rotation Curves}",
    doi = "10.1086/167429",
    journal = "Astrophys. J.",
    volume = "342",
    pages = "635",
    year = "1989"
}

@article{Flanagan:2006ra,
    author = "Flanagan, {\'E}.~{\'E}.",
    title = "{Fourth order Weyl gravity}",
    eprint = "astro-ph/0605504",
    archivePrefix = "arXiv",
    primaryClass = "astro-ph",
    doi = "10.1103/PhysRevD.74.023002",
    journal = "Phys. Rev. D",
    volume = "74",
    pages = "023002",
    year = "2006"
}

@article{Ben09,
    author = "Bengochea, Gabriel R. and Ferraro, Rafael",
    title = "{Dark torsion as the cosmic speed-up}",
    eprint = "0812.1205",
    archivePrefix = "arXiv",
    primaryClass = "astro-ph",
    doi = "10.1103/PhysRevD.79.124019",
    journal = "Phys. Rev. D",
    volume = "79",
    pages = "124019",
    year = "2009"
}

@article{Linder:2010py,
    author = "Linder, Eric V.",
    title = "{Einstein's Other Gravity and the Acceleration of the Universe}",
    eprint = "1005.3039",
    archivePrefix = "arXiv",
    primaryClass = "astro-ph.CO",
    doi = "10.1103/PhysRevD.81.127301",
    journal = "Phys. Rev. D",
    volume = "81",
    pages = "127301",
    year = "2010"
}

@article{Kofinas:2014owa,
    author = "Kofinas, Georgios and Saridakis, Emmanuel N.",
    title = "{Teleparallel equivalent of Gauss-Bonnet gravity and its 
modifications}",
    eprint = "1404.2249",
    archivePrefix = "arXiv",
    primaryClass = "gr-qc",
    doi = "10.1103/PhysRevD.90.084044",
    journal = "Phys. Rev. D",
    volume = "90",
    pages = "084044",
    year = "2014"
}

@article{Bahamonde:2015zma,
    author = {Bahamonde, Sebastian and B{\"o}hmer, Christian G. and Wright, 
Matthew},
    title = "{Modified teleparallel theories of gravity}",
    eprint = "1508.05120",
    archivePrefix = "arXiv",
    primaryClass = "gr-qc",
    doi = "10.1103/PhysRevD.92.104042",
    journal = "Phys. Rev. D",
    volume = "92",
    number = "10",
    pages = "104042",
    year = "2015"
}

@article{Bahamonde:2019shr,
    author = "Bahamonde, Sebastian and Dialektopoulos, Konstantinos F. and Levi 
Said, Jackson",
    title = "{Can Horndeski Theory be recast using Teleparallel Gravity?}",
    eprint = "1904.10791",
    archivePrefix = "arXiv",
    primaryClass = "gr-qc",
    doi = "10.1103/PhysRevD.100.064018",
    journal = "Phys. Rev. D",
    volume = "100",
    number = "6",
    pages = "064018",
    year = "2019"
}

@article{BeltranJimenez:2017tkd,
    author = "Beltr{\'a}n Jim{\'e}nez, Jose and Heisenberg, Lavinia and 
Koivisto, Tomi",
    title = "{Coincident General Relativity}",
    eprint = "1710.03116",
    archivePrefix = "arXiv",
    primaryClass = "gr-qc",
    reportNumber = "NORDITA-2017-100, IFT-UAM/CSIC-17-093, ITS-ETH-2017-10",
    doi = "10.1103/PhysRevD.98.044048",
    journal = "Phys. Rev. D",
    volume = "98",
    number = "4",
    pages = "044048",
    year = "2018"
}

@article{Heisenberg:2023lru,
    author = "Heisenberg, Lavinia",
    title = "{Review on f(Q) gravity}",
    eprint = "2309.15958",
    archivePrefix = "arXiv",
    primaryClass = "gr-qc",
    doi = "10.1016/j.physrep.2024.02.001",
    journal = "Phys. Rept.",
    volume = "1066",
    pages = "1--78",
    year = "2024"
}

@article{De:2023xua,
    author = "De, Avik and Loo, Tee-How and Saridakis, Emmanuel N.",
    title = "{Non-metricity with boundary terms: f(Q,C) gravity and cosmology}",
    eprint = "2308.00652",
    archivePrefix = "arXiv",
    primaryClass = "gr-qc",
    doi = "10.1088/1475-7516/2024/03/050",
    journal = "JCAP",
    volume = "03",
    pages = "050",
    year = "2024"
}

@article{Olive:1989nu,
    author = "Olive, Keith A.",
    title = "{Inflation}",
    doi = "10.1016/0370-1573(90)90074-4",
    journal = "Phys. Rept.",
    volume = "190",
    pages = "307",
    year = "1990"
}

@article{Bartolo:2004if,
    author = "Bartolo, Nicola and Komatsu, Eiichiro and Matarrese, Sabino and 
Riotto, Antonio",
    title = "{Non-Gaussianity from inflation: Theory and observations}",
    eprint = "astro-ph/0406398",
    archivePrefix = "arXiv",
    primaryClass = "astro-ph",
    doi = "10.1016/j.physrep.2004.08.022",
    journal = "Phys. Rept.",
    volume = "402",
    pages = "103",
    year = "2004"
}

@article{Copeland:2006wr,
    author = "Copeland, Edmund J. and Sami, M. and Tsujikawa, Shinji",
    title = "{Dynamics of dark energy}",
    eprint = "hep-th/0603057",
    archivePrefix = "arXiv",
    primaryClass = "hep-th",
    doi = "10.1142/S021827180600942X",
    journal = "Int. J. Mod. Phys. D",
    volume = "15",
    pages = "1753",
    year = "2006"
}

@article{Cai:2009zp,
    author = "Cai, Yi-Fu and Saridakis, Emmanuel N. and Setare, Mohammad R. and 
Xia, Jian-Qiang",
    title = "{Quintom Cosmology: Theoretical implications and observations}",
    eprint = "0909.2776",
    archivePrefix = "arXiv",
    primaryClass = "hep-th",
    doi = "10.1016/j.physrep.2010.04.001",
    journal = "Phys. Rept.",
    volume = "493",
    pages = "1",
    year = "2010"
}

@article{Cai:2006rs,
    author = "Cai, Rong-Gen and Cao, Li-Ming",
    title = "{Unified first law and thermodynamics of apparent horizon in FRW 
Universe}",
    eprint = "gr-qc/0611071",
    archivePrefix = "arXiv",
    primaryClass = "gr-qc",
    doi = "10.1103/PhysRevD.75.064008",
    journal = "Phys. Rev. D",
    volume = "75",
    pages = "064008",
    year = "2007"
}

@article{Renyi:1961EEE,
    author = "R{\'e}nyi, Alfr{\'e}d",
    title = "{On measures of entropy and information}",
    journal = "{\it{Proceedings of the Fourth Berkeley Symposium on 
Mathematical Statistics and Probability}}",
    volume = "1",
    pages = "547--562",
    publisher = "University of California Press",
    year = "1961"
}

@article{Lyra:1998wz,
    author = "Lyra, M. L. and Tsallis, C.",
    title = "{Nonextensivity and multifractality in low dissipative systems}",
    doi = "10.1103/PhysRevLett.80.53",
    journal = "Phys. Rev. Lett.",
    volume = "80",
    pages = "53",
    year = "1998"
}

@article{sharma1975new,
    author = "Sharma, B. D. and Mittal, D. P.",
    title = "{New non-additive measures of entropy for discrete probability 
distributions}",
    journal = "J. Math. Sci.",
    volume = "10",
    pages = "28",
    year = "1975"
}

@article{Kaniadakis:2005zk,
    author = "Kaniadakis, G.",
    title = "{Statistical mechanics in the context of special relativity. II}",
    doi = "10.1103/PhysRevE.72.036108",
    journal = "Phys. Rev. E",
    volume = "72",
    pages = "036108",
    year = "2005"
}

@article{Geng:2019shx,
    author = "Geng, Chao-Qiang and Hsu, Yu-Ting and Lu, Jian-Rong and Yin, 
Ling",
    title = "{Modified Cosmology Models from Thermodynamical Approach}",
    eprint = "1911.06046",
    archivePrefix = "arXiv",
    primaryClass = "astro-ph.CO",
    doi = "10.1140/epjc/s10052-019-7476-y",
    journal = "Eur. Phys. J. C",
    volume = "80",
    number = "1",
    pages = "21",
    year = "2020"
}

@article{Zamora:2022cqz,
    author = "Zamora, D. J. and Tsallis, C.",
    title = "{Thermodynamically consistent entropic late-time cosmological 
acceleration}",
    eprint = "2201.03385",
    archivePrefix = "arXiv",
    primaryClass = "gr-qc",
    doi = "10.1140/epjc/s10052-022-10645-x",
    journal = "Eur. Phys. J. C",
    volume = "82",
    number = "8",
    pages = "689",
    year = "2022"
}

@article{Teimoori:2023hpv,
    author = "Teimoori, Z. and Rezazadeh, K. and Rostami, A.",
    title = "{Inflation based on the Tsallis entropy}",
    eprint = "2307.11437",
    archivePrefix = "arXiv",
    primaryClass = "gr-qc",
    doi = "10.1140/epjc/s10052-024-12435-z",
    journal = "Eur. Phys. J. C",
    volume = "84",
    number = "1",
    pages = "80",
    year = "2024"
}

@article{Naeem:2023ipg,
    author = "Naeem, M. and Bibi, A.",
    title = "{Accelerating universe via entropic models}",
    doi = "10.1140/epjp/s13360-023-04073-3",
    journal = "Eur. Phys. J. Plus",
    volume = "138",
    number = "5",
    pages = "442",
    year = "2023"
}

@article{Jalalzadeh:2023mzw,
    author = "Jalalzadeh, S. and Moradpour, H. and Moniz, P. V.",
    title = "{Modified cosmology from quantum deformed entropy}",
    eprint = "2308.12089",
    archivePrefix = "arXiv",
    primaryClass = "gr-qc",
    doi = "10.1016/j.dark.2023.101320",
    journal = "Phys. Dark Univ.",
    volume = "42",
    pages = "101320",
    year = "2023"
}

@article{Naeem:2023tcu,
    author = "Naeem, M. and Bibi, A.",
    title = "{Correction to the Friedmann equation with Sharma–Mittal entropy: A 
new perspective on cosmology}",
    eprint = "2308.02936",
    archivePrefix = "arXiv",
    primaryClass = "gr-qc",
    doi = "10.1016/j.aop.2024.169618",
    journal = "Annals Phys.",
    volume = "462",
    pages = "169618",
    year = "2024"
}

@article{Coker:2023yxr,
    author = {{\c{C}}oker, Zeynep and {\"O}kc{\"u}, {\"O}zg{\"u}r and Aydiner, 
Ekrem},
    title = "{Modified Friedmann equations from fractional entropy}",
    eprint = "2308.10212",
    archivePrefix = "arXiv",
    primaryClass = "gr-qc",
    doi = "10.1209/0295-5075/acf158",
    journal = "EPL",
    volume = "143",
    number = "5",
    pages = "59001",
    year = "2023"
}

@article{Saavedra:2023rfq,
    author = "Saavedra, J. and Tello-Ortiz, F.",
    title = "{Unified first law of thermodynamics in Gauss–Bonnet gravity on an 
FLRW background}",
    eprint = "2310.08781",
    archivePrefix = "arXiv",
    primaryClass = "gr-qc",
    doi = "10.1140/epjp/s13360-024-05467-7",
    journal = "Eur. Phys. J. Plus",
    volume = "139",
    number = "7",
    pages = "657",
    year = "2024"
}

@article{Nakarachinda:2023jko,
    author = "Nakarachinda, R. and Pongkitivanichkul, C. and Samart, D. and 
Tannukij, L. and Wongjun, P.",
    title = "{Rényi Holographic Dark Energy}",
    eprint = "2312.16901",
    archivePrefix = "arXiv",
    primaryClass = "gr-qc",
    doi = "10.1002/prop.202400073",
    journal = "Fortsch. Phys.",
    volume = "72",
    number = "7-8",
    pages = "2400073",
    year = "2024"
}

@article{Okcu:2024tnw,
    author = {{\"O}kc{\"u}, {\"O}zg{\"u}r},
    title = "{Investigation of generalised uncertainty principle effects on FRW 
cosmology}",
    eprint = "2401.09477",
    archivePrefix = "arXiv",
    primaryClass = "gr-qc",
    doi = "10.1016/j.nuclphysb.2024.116551",
    journal = "Nucl. Phys. B",
    volume = "1004",
    pages = "116551",
    year = "2024"
}

@article{Jalalzadeh:2024qej,
    author = "Jalalzadeh, R. and Jalalzadeh, S. and Jahromi, A. S. and 
Moradpour, H.",
    title = "{Friedmann equations of the fractal apparent horizon}",
    eprint = "2404.06986",
    archivePrefix = "arXiv",
    primaryClass = "gr-qc",
    doi = "10.1016/j.dark.2024.101498",
    journal = "Phys. Dark Univ.",
    volume = "44",
    pages = "101498",
    year = "2024"
}

@article{Trivedi:2024inb,
    author = "Trivedi, O. and Bidlan, A. and Moniz, P.",
    title = "{Fractional holographic dark energy}",
    eprint = "2407.16685",
    archivePrefix = "arXiv",
    primaryClass = "gr-qc",
    doi = "10.1016/j.physletb.2024.139074",
    journal = "Phys. Lett. B",
    volume = "858",
    pages = "139074",
    year = "2024"
}

@article{Okcu:2024llu,
    author = {{\"O}kc{\"u}, {\"O}zg{\"u}r and Aydiner, Ekrem},
    title = "{Exponential correction to Friedmann equations}",
    eprint = "2407.14685",
    archivePrefix = "arXiv",
    primaryClass = "gr-qc",
    doi = "10.1007/s10714-024-03273-1",
    journal = "Gen. Rel. Grav.",
    volume = "56",
    number = "7",
    pages = "87",
    year = "2024"
}

@article{Ens:2024zzs,
    author = "Ens, P. S. and Santos, A. F.",
    title = "{Barrow holographic dark energy: a path to reconstructing f(R,T) 
gravity}",
    eprint = "2412.09189",
    archivePrefix = "arXiv",
    primaryClass = "gr-qc",
    doi = "10.1140/epjc/s10052-024-13708-3",
    journal = "Eur. Phys. J. C",
    volume = "84",
    number = "12",
    pages = "1338",
    year = "2024"
}

@article{Ualikhanova:2024xxe,
    author = "Ualikhanova, Ulbossyn and Altaibayeva, Aziza and Chattopadhyay, 
Surajit",
    title = "{Holographic reconstruction of k-essence model with Tsallis and the 
most generalized Nojiri-Odintsov version of holographic dark energy}",
    eprint = "2501.04028",
    archivePrefix = "arXiv",
    primaryClass = "physics.gen-ph",
    doi = "10.1007/s12648-025-03606-z",
    journal = "Indian J. Phys.",
    volume = "99",
    number = "11",
    pages = "4433--4441",
    year = "2025"
}

@article{Shahhoseini:2025sgl,
    author = "Shahhoseini, N. and Malekjani, M. and Khodam-Mohammadi, A.",
    title = "{$\varLambda $CDM model against gravity-thermodynamics conjecture: observational constraints after DESI 2024}",
    eprint = "2501.03655",
    archivePrefix = "arXiv",
    primaryClass = "astro-ph.CO",
    doi = "10.1140/epjc/s10052-025-13781-2",
    journal = "Eur. Phys. J. C",
    volume = "85",
    number = "1",
    pages = "53",
    year = "2025"
}

@article{Nojiri:2025gkq,
    author = "Nojiri, Shin'ichi and Odintsov, Sergei D. and Paul, Tanmoy and 
SenGupta, Soumitra",
    title = "{Modified gravity as entropic cosmology}",
    eprint = "2503.19056",
    archivePrefix = "arXiv",
    primaryClass = "gr-qc",
    month = "3",
    year = "2025"
}

@article{Nojiri:2022dkr,
    author = "Nojiri, Shin'ichi and Odintsov, Sergei D. and Paul, Tanmoy",
    title = "{Early and late universe holographic cosmology from a new 
generalized entropy}",
    eprint = "2205.08876",
    archivePrefix = "arXiv",
    primaryClass = "gr-qc",
    doi = "10.1016/j.physletb.2022.137189",
    journal = "Phys. Lett. B",
    volume = "831",
    pages = "137189",
    year = "2022"
}

@article{Moresco:2012by,
    author = "Moresco, Michele and Verde, Licia and Pozzetti, Lucia and Jimenez, 
Raul and Cimatti, Andrea",
    title = "{New constraints on cosmological parameters and neutrino properties 
using the expansion rate of the Universe to z{\textasciitilde}1.75}",
    eprint = "1201.6658",
    archivePrefix = "arXiv",
    primaryClass = "astro-ph.CO",
    doi = "10.1088/1475-7516/2012/07/053",
    journal = "JCAP",
    volume = "07",
    pages = "053",
    year = "2012"
}

@article{Moresco:2015cya,
    author = "Moresco, Michele",
    title = "{Raising the bar: new constraints on the Hubble parameter with 
cosmic chronometers at z {\ensuremath{\sim}} 2}",
    eprint = "1503.01116",
    archivePrefix = "arXiv",
    primaryClass = "astro-ph.CO",
    doi = "10.1093/mnrasl/slv037",
    journal = "Mon. Not. Roy. Astron. Soc.",
    volume = "450",
    number = "1",
    pages = "L16--L20",
    year = "2015"
}

@article{Moresco:2016mzx,
    author = "Moresco, Michele and Pozzetti, Lucia and Cimatti, Andrea and 
Jimenez, Raul and Maraston, Claudia and Verde, Licia and Thomas, Daniel and 
Citro, Annalisa and Tojeiro, Rita and Wilkinson, David",
    title = "{A 6{\%} measurement of the Hubble parameter at $z\sim0.45$: direct 
evidence of the epoch of cosmic re-acceleration}",
    eprint = "1601.01701",
    archivePrefix = "arXiv",
    primaryClass = "astro-ph.CO",
    doi = "10.1088/1475-7516/2016/05/014",
    journal = "JCAP",
    volume = "05",
    pages = "014",
    year = "2016"
}

@article{Scolnic:2021amr,
    author = "Scolnic, Dan and others",
    title = "{The Pantheon+ Analysis: The Full Data Set and Light-curve 
Release}",
    eprint = "2112.03863",
    archivePrefix = "arXiv",
    primaryClass = "astro-ph.CO",
    doi = "10.3847/1538-4357/ac8b7a",
    journal = "Astrophys. J.",
    volume = "938",
    number = "2",
    pages = "113",
    year = "2022"
}

@article{DESI:2025zgx,
    author = "Abdul Karim, M. and others",
    collaboration = "DESI",
    title = "{DESI DR2 results. II. Measurements of baryon acoustic oscillations 
and cosmological constraints}",
    eprint = "2503.14738",
    archivePrefix = "arXiv",
    primaryClass = "astro-ph.CO",
    reportNumber = "FERMILAB-PUB-25-0169-PPD",
    doi = "10.1103/tr6y-kpc6",
    journal = "Phys. Rev. D",
    volume = "112",
    number = "8",
    pages = "083515",
    year = "2025"
}

@article{Brout:2022vxf,
    author = "Brout, Dillon and others",
    title = "{The Pantheon+ Analysis: Cosmological Constraints}",
    eprint = "2202.04077",
    archivePrefix = "arXiv",
    primaryClass = "astro-ph.CO",
    doi = "10.3847/1538-4357/ac8e04",
    journal = "Astrophys. J.",
    volume = "938",
    number = "2",
    pages = "110",
    year = "2022"
}

@ARTICLE{2020ApJ...898...82M,
       author = {{Moresco}, Michele and {Jimenez}, Raul and {Verde}, Licia and {Cimatti}, Andrea and {Pozzetti}, Lucia},
        title = "{Setting the Stage for Cosmic Chronometers. II. Impact of Stellar Population Synthesis Models Systematics and Full Covariance Matrix}",
      journal = {\apj},
         year = 2020,
        month = jul,
       volume = {898},
       number = {1},
          eid = {82},
        pages = {82},
          doi = {10.3847/1538-4357/ab9eb0},
archivePrefix = {arXiv},
       eprint = {2003.07362},
 primaryClass = {astro-ph.GA},
       adsurl = {https://ui.adsabs.harvard.edu/abs/2020ApJ...898...82M}
}

@article{Chantada2023,
  title = {Cosmology-informed neural networks to solve the background dynamics of the {Universe}},
  author = {Chantada, Augusto T. and Landau, Susana J. and Protopapas, Pavlos and Sc\'occola, Claudia G. and Garraffo, Cecilia},
  journal = {Phys. Rev. D},
  volume = {107},
  issue = {6},
  pages = {063523},
  numpages = {28},
  year = {2023},
  month = {Mar},
  publisher = {American Physical Society},
  doi = {10.1103/PhysRevD.107.063523}
}

@ARTICLE{2024PhLB..85438717L,
       author = {{Liu}, Guanlin and {Wang}, Yu and {Zhao}, Wen},
        title = "{Testing the consistency of early and late cosmological parameters with BAO and CMB data}",
      journal = {Physics Letters B},
         year = 2024,
        month = jul,
       volume = {854},
          eid = {138717},
        pages = {138717},
          doi = {10.1016/j.physletb.2024.138717},
archivePrefix = {arXiv},
       eprint = {2401.10571},
 primaryClass = {astro-ph.CO},
       adsurl = {https://ui.adsabs.harvard.edu/abs/2024PhLB..85438717L}
}

@article{Torrado:2020dgo,
    author = "Torrado, Jesus and Lewis, Antony",
    title = "{Cobaya: Code for Bayesian Analysis of hierarchical physical models}",
    eprint = "2005.05290",
    archivePrefix = "arXiv",
    primaryClass = "astro-ph.IM",
    reportNumber = "TTK-20-15",
    doi = "10.1088/1475-7516/2021/05/057",
    journal = "JCAP",
    volume = "05",
    pages = "057",
    year = "2021"
}

@misc{2019ascl.soft10019T,
       author = {{Torrado}, Jes{\'u}s and {Lewis}, Antony},
        title = "{Cobaya: Bayesian analysis in cosmology}",
 howpublished = {Astrophysics Source Code Library, record ascl:1910.019},
         year = 2019,
        month = oct,
          eid = {ascl:1910.019},
archivePrefix = {ascl},
       eprint = {1910.019},
       adsurl = {https://ui.adsabs.harvard.edu/abs/2019ascl.soft10019T}
}

@ARTICLE{2011JCAP...07..034B,
       author = {{Blas}, Diego and {Lesgourgues}, Julien and {Tram}, Thomas},
        title = "{The Cosmic Linear Anisotropy Solving System (CLASS). Part II: Approximation schemes}",
      journal = {\jcap},
         year = 2011,
        month = jul,
       volume = {2011},
       number = {7},
          eid = {034},
        pages = {034},
          doi = {10.1088/1475-7516/2011/07/034},
archivePrefix = {arXiv},
       eprint = {1104.2933},
 primaryClass = {astro-ph.CO},
       adsurl = {https://ui.adsabs.harvard.edu/abs/2011JCAP...07..034B}
}

@article{Luciano:2025hjn,
    author = "Luciano, Giuseppe Gaetano and Paliathanasis, Andronikos and Saridakis, Emmanuel N.",
    title = "{Barrow and Tsallis entropies after the DESI DR2 BAO data}",
    eprint = "2504.12205",
    archivePrefix = "arXiv",
    primaryClass = "gr-qc",
    doi = "10.1088/1475-7516/2025/09/013",
    journal = "JCAP",
    volume = "09",
    pages = "013",
    year = "2025"
}



\end{document}